\documentclass{article}

\newcommand{\Cc}[1]{{\mathcal C}_{#1}}

\usepackage[english]{babel}
\usepackage{cancel}
\usepackage[letterpaper,top=2cm,bottom=2cm,left=3cm,right=3cm,marginparwidth=1.75cm]{geometry}

\usepackage{amsmath}
\usepackage{graphicx}
\usepackage[colorlinks=true, allcolors=blue]{hyperref}
\usepackage{xcolor}

\usepackage{authblk}
\usepackage{etoolbox}

\makeatletter

\patchcmd{\@affil}{\centering}{\raggedright}{}{}

\makeatother
\title{Probing unknown nonperturbative effects in $b \to s \ell\ell$ with inclusive and exclusive observables}
\author[1]{P. Alvarez-Cartelle\thanks{Email: \texttt{paula.alvarez@usc.es}}}
\affil[1]{\footnotesize Instituto Galego de F\'isica de Altas Enerx\'ias (IGFAE), Universidade de Santiago de Compostela,
Santiago de Compostela, Spain}
\affil[1]{Cavendish Laboratory, University of Cambridge, Cambridge, United Kingdom}
\author[2]{B. Capdevila\thanks{Email: \texttt{bcapdevila@icc.ub.edu}}}
\affil[2]{Departament de F\'isica Qu\'antica i Astrof\'isica, Institut de Ci\`encies del Cosmos (ICCUB), Universitat de Barcelona, Mart\'i Franqu\`es 1, E08028 Barcelona}
\author[3]{E. Lunghi\thanks{Email: \texttt{elunghi@iu.edu}}}
\affil[3]{Physics Department, Indiana University, Bloomington, IN 47405, USA}
\author[4]{J. Matias\thanks{Email: \texttt{matias@ifae.es}}}
\affil[4]{Universitat Aut\`onoma de Barcelona, 08193 Bellaterra, Barcelona}
\affil[4]{Institut de F\'isica d’Altes Energies (IFAE), The Barcelona Institute of Science and Technology, Campus
UAB, 08193 Bellaterra (Barcelona)}

\begin{document}
\maketitle

\begin{abstract}
In this paper we revisit, from a different perspective, a long-standing question: ``Is the systematic deficit observed in all branching ratios mediated by a $b\to s \mu\mu$ transition due to New Physics, or to a hypothetical constant unknown universal hadronic contribution that mimics New Physics?'' The key observation that allows us to distinguish between these two possibilities is that non-perturbative contributions associated with $c\bar c$ loops affect inclusive $B\to X_s \ell\ell$ and exclusive $B\to K^{(*)}\ell\ell$ modes differently. In inclusive decays, factorizable contributions are exactly determined from data on $e^+e^-\to \mathrm{hadrons}$, while non-factorizable corrections are described by resolved-photon contributions at low $q^2$ and by local power corrections at high $q^2$. In exclusive decays, by contrast, hypothetical charming-penguin effects, beyond those already included in current uncertainty estimates, could appear, in a worst-case scenario, as a constant, universal contribution that it seems, in principle, indistinguishable from genuine New Physics. We identify two observables, constructed from ratios of exclusive to inclusive modes, that can discriminate between a New Physics contribution and a constant hadronic contribution. Moreover, these ratios can be measured directly by LHCb, as they do not require any normalisation to $J/\psi K^{(*)}$ branching fractions from B factories. A preliminary evaluation of these observables with present data shows some preference for the New Physics interpretation. In a complementary test, a comparison between inclusive measurements and the corresponding sum of exclusive modes at high $q^2$ similarly disfavours an explanation based on a constant hadronic contribution. Finally, we provide projections for the new observables based on expected LHCb and Belle II measurements in the near future.
\end{abstract}

\section{Introduction and Motivation}\label{sec:intro}
Tensions between experimental measurements and Standard Model (SM) predictions for several $B$-meson decays, so called, $B$-flavour anomalies not only remain but have been further reinforced with very recent updates from LHCb~\cite{LHCb:2025mqb} and CMS~\cite{CMS:2024atz}. 

On the theory side, factorizable QCD effects are absorbed into local hadronic form factors, for which several determinations exist using both continuum approaches, most prominently Light-Cone Sum Rules (LCSRs), and lattice QCD computations~\cite{Khodjamirian:2010vf,Gubernari:2018wyi,Bharucha:2015bzk,Horgan:2015vla,Parrott:2022rgu}. Non-local QCD effects are typically addressed through a combination of QCD factorization (QCDf) and phenomenological modelling of the next-to-leading-order (NLO) soft-gluon correction to the non-local form factor~\cite{Ciuchini:2022wbq,Alguero:2023jeh,Hurth:2025vfx}. Existing estimates of the soft-gluon contribution either rely on computations in the light-cone operator product expansion (LCOPE) at small and negative $q^2$ (with $q^2$ denoting the dilepton invariant mass squared) and on dispersive relations to continue these results to the physical region, or they adopt phenomenological ans\"atze. More recently, approaches based on extended LCOPE calculations for the non-local form factor, combined with a $z$-parameterization of its analytic structure and unitarity constraints imposed via dispersion relations, have also become standard tools in the field~\cite{Gubernari:2022hxn,Gubernari:2023puw}.

Within these different theoretical frameworks, several groups perform global fits to $b\to s\ell^+\ell^-$ data~\cite{Alguero:2023jeh,Hurth:2025vfx,Gubernari:2023puw}, typically including datasets of $\mathcal{O}(200)$ observables or more. Despite differences in the treatment of hadronic effects and in the selection of experimental inputs, these analyses consistently find a tension between Standard Model (SM) predictions and the measured data, with the quoted statistical significance varying across fits. Among the most prominent and persistent discrepancies are those observed in the central $q^2$ bins of the angular observable $P_5^\prime$~\cite{Matias:2012xw,Descotes-Genon:2013vna}, as well as in the branching ratio of the $B\to K \mu\mu$ decay~\cite{Alguero:2023jeh,Parrott:2022zte}.

Ref.~\cite{Ciuchini:2022wbq} suggested that charm rescattering, involving $D_s$, $D^{(*)}$ intermediate states, could generate an additional contribution to exclusive $b\to s\ell^+\ell^-$ amplitudes that enters in the same way as an effective shift in $\mathcal{C}_9$, the Wilson coefficient of the semileptonic operator,
\begin{equation}\label{eq:erO9}
    O_{9\ell}=\frac{e^2}{16 \pi^2} \Big(\bar{s} \gamma_\mu P_L b\Big) \Big(\bar{\ell} \gamma^\mu \ell\Big).
\end{equation}
This contribution was first estimated for the $B\to K\mu^+\mu^-$ channel in Refs.~\cite{Isidori:2024lng,Isidori:2025dkp}, based on Heavy-hadron Chiral Perturbation Theory (HHChPT), an extrapolation prescription to cover the full kinematic region, and phenomenological assumptions to combine the various intermediate modes. Two qualitative features emerge. 
First, this mechanism has difficulties to reach the size of some of the observed anomalies even under the extreme assumption of fully constructive interference among all modes. 
Second and more importantly, monopole and dipole photon-hadron couplings yield contributions that have opposite and same signs at low- and high-$q^2$, respectively. This implies that if all contributions add constructively at low-$q^2$, some cancellations are expected at high-$q^2$, yielding a smaller overall effect.  
This observation, combined with the strong $q^2$ dependence of the HHChPT estimate, suggests a violation of universality that is in tension with global-fit results (including bin-by-bin analyses), which show a clear preference for a single universal contribution to $\mathcal{C}_9$.

One has to acknowledge, at this point, that the low-$q^2$ region in Refs.~\cite{Isidori:2024lng,Isidori:2025dkp} may carry sizeable systematic uncertainties due to the extrapolation strategy from the high-$q^2$ estimate. Moreover, it is not yet fully settled whether this rescattering contribution is already included, in whole or in part, in the NLO calculation of Ref.~\cite{Asatrian:2019kbk}.

On the experimental side, the branching ratios of the channels $B^{(0,+)}\to K^{(0,+)}\mu\mu$, $B^{(0,+)}\to K^{*(0,+)}\mu^+\mu^-$ and $B_s\to\phi\mu^+\mu^-$ show a consistent deficit in data with respect to SM predictions, both at the low and  high-$q^2$  regions\footnote{Low- and high-$q^2$ denote regions below and above the two narrow charmonium resonances $J/\psi$ and $\psi(2S)$. The precise ranges used in experimental analyses depend on the decay mode and on the adopted binning.}. At high-$q^2$ it is customary to use a wide bin in order to smooth the impact of broad resonances. In the approach of Ref.~\cite{Alguero:2023jeh}, which we take as the default framework for exclusive $b\to s\ell^+\ell^-$ modes, a conservative flat uncertainty of order $10\%$ is applied at large $q^2$ on top of other theory errors, in order to account for potential quark-hadron duality violations\footnote{In Ref.\cite{Beylich:2011aq}  these quark-hadron duality violations were estimated to be of order 3\%. }. The observed pattern of deficits across channels, as it can be concluded from the various global fits, can be accommodated by a universal NP contribution, which we denote by $C_9^{\rm U}$. 

The universality of such a NP interpretation has been scrutinized in dedicated studies. In Ref.~\cite{Alguero:2023jeh}, the variation of $C_9^{\rm U}$ across different $q^2$ regions, bins in $q^2$ and across the various processes entering the global fit was tested, with results compatible with universality at the $68\%$ confidence level. Ref.~\cite{Bordone:2024hui} performed a similar analysis using a resonance-model estimate of the non-local form factor and found consistent conclusions, although focusing on $B\to K\ell^+\ell^-$ data only. This observed universality would be challenging to replicate with a rescattering contribution of the type advocated in Ref.~\cite{Ciuchini:2022wbq} if 
the latter follows the sign pattern at low- and high-$q^2$ suggested by the results in Refs.~\cite{Isidori:2024lng,Isidori:2025dkp}.

Taken together, the previous considerations point to three logical possibilities:

\begin{itemize}
\item[\textit{i})] 
The rescattering contribution is already included in the NLO calculation of Ref.~\cite{Asatrian:2019kbk} and should not be considered to avoid double counting. A recent analysis, see Ref.~\cite{Hoferichter:2026jlh}, demonstrated that the hadronic anomalous thresholds discussed in Ref.~\cite{Ciuchini:2022wbq} have an exact counterpart in the corresponding quark-level diagrams. Further work along these lines might be able to clarify the extent to which partonic calculations already capture the entirety of these effects.

\item[\textit{ii})] 
The calculation of Refs.~\cite{Isidori:2024lng, Isidori:2025dkp} can be trusted to describe the relative signs and the strong $q^2$ dependence of monopole and dipole contributions at low- and high-$q^2$. Therefore this hadronic contribution is actually disfavored as a solution of the anomalies by the global fits that point to a deficit in several branching ratios and that are compatible with a flat contribution to ${\cal C}_9$.

\item[\textit{iii})] Other hypothetical unknown extra hadronic effects, possibly including the rescattering contributions of Ref.~\cite{Ciuchini:2022wbq} but not necessarily limited to it, could conspire to generate an effective contribution that resembles a universal shift in $\mathcal{C}_9$. In the following, we denote this additional hypothetical term by ${\cal C}_{9,\rm extra}^{K^{(*)},\rm low \, or\, high}(q^2)$ for $B\to K$ and $B\to K^*$ transitions (at low and high-$q^2$), respectively. Note however, that universality tests currently still lack the precision required to unambiguously resolve a non-trivial $q^2$ dependence. Finally, although it appears unlikely, one could contemplate issues with the experimental measurements.
\end{itemize}

\noindent While experimental and theoretical improvements are ongoing, in this paper we propose a different approach to constrain the size of a hypothetical hadronic contribution that mimics the structure of a universal NP shift in $\mathcal{C}_9$. Guided by the discussion above, we consider three possible scenarios:

\begin{itemize}
\item[A:] \textit{Universal case}. The hypothetical contribution is constant in $q^2$ and universal across channels. It cannot  arise from the rescattering processes estimated in Refs.~\cite{Isidori:2024lng,Isidori:2025dkp} because of the sign 
correlations between low- and high-$q^2$. Instead, it can arise from some other unknown constant hadronic source, that imitates a universal NP contribution. In this scenario one can combine different channels and $q^2$ regions to test the universality assumption directly.

\item[B:] \textit{Universality among channels}. The contribution is universal across channels but not constant in $q^2$, for instance it could differ at low and high-$q^2$. This is in tension with the fact that the deviations in branching ratios display the same sign at low and high-$q^2$ in several modes, including $B\to K\mu\mu$.

\item[C:] \textit{Non-universal case}. The contribution is neither universal in $q^2$ nor universal across channels. In this case it cannot mimic the NP pattern preferred by current global fits and would typically imply an observable mode dependence and/or a pronounced $q^2$ dependence, neither of which is clearly supported by data at present.
\end{itemize}

\noindent Scenarios B and C are disfavored by current data and, more importantly, their viability will be tested more and more as data accumulates because they predict a non-trivial pattern across channels and kinematic regions. The only scenario that is not immediately in conflict with the observed coherence of the deviations is scenario A: a universal constant hadronic contribution that would imitate a universal NP shift. This is therefore the relevant worst-case scenario, and the one on which we focus.

The main goal of this paper is to propose a test that can rule out this worst-case possibility. Our key observation is that ratios involving inclusive and exclusive $b\to s\ell\ell$ branching fractions can discriminate between a genuine NP contribution to $\mathcal{C}_9$ and a hypothetical constant hadronic contribution: exclusive and inclusive modes are expected to display different sensitivity to charm rescattering effects. Moreover, the observables we construct do not suffer from the normalization limitations that affect absolute branching ratios, which should translate into a significantly improved experimental precision. Other scenarios, like the possible appearance of large strong phases as proposed in Ref.~\cite{Altmannshofer:2026cwk}, can be likely tested with the same methodology discussed in this paper.

The use of inclusive measurements at low- and high-$q^2$ to confirm the NP interpretation of anomalies in the various exclusive channels is not novel. This has been done over the years (see, for instance, Refs.~\cite{Huber:2020vup}) by focusing mostly on the low-$q^2$ region. More recently the analyses in Refs.~\cite{Isidori:2023unk, Huber:2024rbw} stressed that possibility of reconstructing the inclusive high-$q^2$ branching ratio using LHCb data only and brought in focus the importance of the high-$q^2$ region. In this work, we sharpen this approach
but our focus is not exclusively on NP but on probing the absence of unknown hadronic contributions that mimic NP and on the construction of observables that can be measured at LHCb without need for {\it any} external normalizations.

The paper is organized as follows. In Section~\ref{sec:excinc} we discuss the theoretical description of the exclusive $B\to K\ell\ell$ and $B\to K^*\ell\ell$ decays compared with the inclusive $B\to X_s\ell\ell$ mode, with particular emphasis on the absence of the hypothetical hadronic contribution in the inclusive case. We also discuss the structure of the Wilson coefficient $\mathcal{C}_9$ in exclusive and inclusive observables. In Section~\ref{sec:observables} we define two new observables that implement our test and provide their dependence on $\mathcal{C}_9$ for the relevant modes. In Section~\ref{sec:F_exp} we use available data to obtain first experimental determinations of these observables in a given $q^2$ region. In Section~\ref{sec:current} we explore the different behavior of the new observables as a function of ${\cal C}_{9,\rm extra}$ or ${\cal C}_9^{\rm NP}$ and describe the possible directions of improvement in the near future. In Section~\ref{sec:inclusive-exclusive} we compare the inclusive computation with the sum of exclusive channels to constrain a hypothetical extra constant contribution. We conclude in Section~\ref{sec:conclusions}.

\section{Theoretical Framework: Exclusive versus inclusive}
\label{sec:excinc}
We begin with some introductory remarks on the four distinct theoretical frameworks required to describe inclusive and exclusive $b\to s\ell\ell$ decays at low and high-$q^2$.

The weak effective Hamiltonian relevant to the processes considered here is~\cite{Buchalla:1995vs,Grinstein:1987vj}:
\begin{equation}
{\mathcal H}_{\rm eff}=-\frac{4G_F}{\sqrt{2}} V_{tb}V_{ts}^*\sum_i {\mathcal C}_i  {\mathcal O}_i + {\rm h.c.}
\end{equation}
See Ref.~\cite{Capdevila:2017bsm} for explicit definitions of the relevant electromagnetic and semileptonic operators beyond ${\cal O}_{9\ell}$, introduced above in Eq.~\eqref{eq:erO9}. These other operators will not play a significant role in the present discussion, since we restrict ourselves to NP contributions to $\mathcal{C}_9$. This simplified setup is motivated by the outcome of global fits.

\subsection{Exclusive at low and high-\texorpdfstring{$q^2$}{q2}}\label{subsec:exclusive}

\underline{Exclusive decays at low-$q^2$} are described within the SCET-II framework. The resulting leading power amplitudes are given in terms of form factors and meson light-cone distribution amplitudes. Power corrections are nonlocal and can be formally written in terms of unknown subleading form factors and higher twist distribution amplitudes (see, for instance, the discussions in Refs.~\cite{Hardmeier:2003ig, Becher:2003qh}). This lack of knowledge of the nonperturbative inputs required to estimate these nonlocal effects (commonly referred to as ``charming penguins'') yields uncertainties whose size is the subject of an animated debate in the literature. 

In line with the analyses of Refs.~\cite{Descotes-Genon:2015uva,Capdevila:2017bsm,Alguero:2019ptt,Alguero:2023jeh}, we adopt the phenomenological model of Ref.~\cite{Khodjamirian:2010vf} and include a $q^2$, polarisation and process-dependent non-local contribution to $\Cc{9\ell}^{\rm eff}$
\begin{equation}
\label{decomp}
{\cal C}_{9\ell}^{{\rm eff}} \to {\cal C}_{9\ell}^{{\rm eff}} = {\cal C}_{9\ell,\,\rm pert}^{\rm SM} + {\cal C}_{9\ell}^{\rm NP} + {\cal C}_{9,\,j}^{c\bar{c}\, B \to K^{(*)}}  
\end{equation}
where $j=\perp,\parallel,0$ and
\begin{equation}
    {\cal C}_{9\ell,\,\rm pert}^{\rm SM}=C_{9\ell}^{\rm SM}+Y(q^2)
\end{equation}
where the definition of $Y(q^2)$ can be found in Ref.~\cite{Beneke:2001at}, while the corresponding $\alpha_s$ corrections are given in Ref.~\cite{Asatrian:2019kbk}. Finally, $\mathcal{C}_{9,\, j}^{c\bar{c}\, B \to K^{(*)}} = s_j\,\mathcal{C}_{9,\,j\,\rm KMPW}^{c\bar{c}\, B \to K^{(*)}}$ denotes the phenomenological model for the soft-gluon non-factorizable contribution computed in Ref.~\cite{Khodjamirian:2010vf} and that is included in the SM predictions of Refs.~\cite{Descotes-Genon:2015uva,Capdevila:2017bsm,Alguero:2019ptt,Alguero:2023jeh}. In the $B\to K^*\ell\ell$ case, this contribution depends on the polarisation of the underlying transversity amplitude denoted by $j$. Since the analysis of Ref.~\cite{Khodjamirian:2010vf} did not include the imaginary part of the corresponding non-local form factor, the parameters $s_j$ were introduced in Ref.~\cite{Descotes-Genon:2015uva} and allowed to vary in the interval $[-1,1]$ in order to account for the uncertainty associated with the relative phase between the short-distance and long-distance contributions to $\mathcal{C}_9$.

In recent years, significant theoretical progress has been made in evaluating the long-distance charm contribution entering ${\cal C}_9^{\rm eff}$, either through explicit computations or by using resonance data to model it~\cite{Bobeth:2017vxj}. More recently, Ref.~\cite{Gubernari:2020eft} presented a more complete calculation, including the imaginary part and higher-twist corrections. That analysis confirmed the qualitative picture obtained in Ref.~\cite{Khodjamirian:2010vf}, while improving the precision through updated inputs and a more refined treatment. Importantly, Ref.~\cite{Gubernari:2020eft} found that the numerical impact of the soft-gluon contribution to the non-local form factor is reduced when including all light-cone distribution amplitudes. This result is an important milestone in quantifying the impact of charm contributions in this decay. By contrast, this contribution is not included in $B\to K\mu^+\mu^-$ decays, since both Refs.~\cite{Khodjamirian:2010vf,Gubernari:2020eft} find it to be numerically negligible for this channel.

According to the decomposition in Eq.~\eqref{decomp}, any additional hypothetical hadronic contribution, arising for instance from higher-order non-perturbative charm effects or from some other unknown constant hadronic source, would enter as an extra term on top of ${\cal C}_9^{\rm eff}(q^2)$ in Eq.~\eqref{decomp}. Such a contribution would therefore compete with the long-distance charm component already included in our analysis, for which we adopt a deliberately conservative treatment. This additional hypothetical term corresponds to what we denoted in the introduction as  ${\cal C}_{9,\rm extra}^{K^{(*)},\rm low}(q^2)$ for $B\to K$ and $B\to K^*$ transitions (at low-$q^2$), respectively. Since in this work we focus on the worst-case scenario A, namely a constant hadronic contribution capable of mimicking NP, we will mainly focus our attention on the case in which this term is process, helicity and $q^2$ independent,
\begin{equation}
{\cal C}_{9,\rm extra}^{K^{(*)},\rm high}= {\cal C}_{9,\rm extra}^{K^{(*)},\rm low}={\cal C}_{9,\rm extra} 
\end{equation}
For the local form factors, at low-$q^2$ we use LCSR only inputs. At high-$q^2$ we rely on lattice-QCD determinations, treating the two kinematic regions independently through their corresponding $z$-parameterizations.
\medskip

\noindent\underline{Exclusive decays at high-$q^2$} are usually discussed in the context of an expansion in $1/q^2$~\cite{Grinstein:2004vb}. The resulting amplitude is thus expressed in terms of form factors and local power corrections only. If this large $q^2$ expansion is interpreted as an OPE, $q^\mu$ is the conjugated momentum associated with the spacetime separation between the insertion of the operators ${\mathcal O}_{1,2}$ and the electromagnetic current (which produces the two leptons); thus, the charmonium resonances are precisely the intermediate states which are included via quark-hadron duality. This shows that the Kr\"uger-Sehgal (KS) approach of Ref.~\cite{Kruger:1996cv} in this situation is meaningless (it would be akin to trying to model resonances in $e^+e^- \to {\rm hadrons}$ while having at hand the OPE). In the low-recoil region, the corresponding ${\cal C}_{9\ell}^{\rm eff}$ is given by~\cite{Bobeth:2010wg}
\begin{eqnarray} \label{decomp2}
{\cal C}_{9\ell}^{\rm eff}(q^2)&=&\!\!\!\!{\cal C}_{9\ell}^{\rm SM}+{\cal C}_{9\ell}^{\rm NP}+ h(0,q^2)\left[\frac{4}{3} {\cal C}_1+{\cal C}_2+\frac{11}{2}{\cal C}_3-\frac{2}{3} {\cal C}_4+ 52 {\cal C}_5 - \frac{32}{3}{\cal C}_6\right] \hfill\nonumber \\ &-&\!\!\!\!\frac{1}{2}h(m_b,q^2) \left[7 {\cal C}_3+\frac{4}{3} {\cal C}_4+76 {\cal C}_5+\frac{64}{3}{\cal C}_6\right]+\frac{4}{3}\left[ {\cal C}_3+\frac{16}{3}{\cal C}_5 + \frac{16}{9}{\cal C}_6\right] \nonumber \\
&+&\!\!\!\!8\frac{m_c^2}{q^2} \left[\frac{4}{9}{\cal C}_1+\frac{1}{3}{\cal C}_2+ 2 {\cal C}_3+20{\cal C}_5\right] \hfill
\end{eqnarray}
where 4-quark operators are included. This expression receives $\alpha_s$ corrections that can be found in Ref.~\cite{Bobeth:2010wg}. See the same reference for the definition of $h(m_q, q^2)$. Again here in addition one would have to consider any hypothetical extra hadronic contribution denoted by ${\cal C}_{9,\rm extra}^{K^{(*)},\rm high}(q^2)$.

This completes all the possible SM contributions, but of course, one also has  the NP contribution to the Wilson coefficient, ${\cal C}_{9\ell}^{\rm NP}$ in Eq.(\ref{decomp}) and Eq.(\ref{decomp2}). From the global analysis of more than 200  observables in Ref.\cite{Alguero:2023jeh}, {there is a clear preference for a lepton-flavour universal NP contribution, i.e., $\mathcal{C}_{9\mu}^\mathrm{NP} = \mathcal{C}_{9e}^\mathrm{NP} = \mathcal{C}_9^{\rm NP}$.} 

For the local form factors entering the high-$q^2$ exclusive amplitudes, we use lattice-QCD inputs fitted through an appropriate $z$-expansion. No combined fit with the low-$q^2$ determinations is performed.

\subsection{Inclusive at low and high-\texorpdfstring{$q^2$}{q2}}\label{subsec:inclusive}

\underline{Inclusive decays at low-$q^2$} are described in terms of an Operator Product Expansion (OPE). Note that the momentum conjugated to the coordinate separation between the insertion of any two operators is the four-momentum of the $X_s$ state. In this sense, the sum over all hadronic states with one unit of strangeness is described, modulo quark-hadron duality violating effects, by the OPE itself. This picture fails when the operators ${\mathcal O}_{1,2}$, which contain a pair of charm fields, are considered. The requirement to include only cuts through the two leptons, while neglecting purely hadronic ones, exposes the OPE to contributions which diverge in the limit in which the widths of the narrow resonances vanish~\cite{Beneke:2009az}. These resonant effects stem from diagrams which, at the quark level, involve a charm loop.

In order to address this issue, it is convenient to separate factorizable and nonfactorizable charm loop diagrams. 
The former can be exactly described in terms of the $e^+e^- \to {\rm hadrons}$ data using a dispersion relation (KS mechanism). Intuitively, these factorizable effects yield an OPE which is local in the $X_s$ separation but nonlocal in the separation between the insertion of the two operators and the emission of the virtual photon. The replacement of the perturbative two loop charm contribution to the OPE with the KS ansatz yields a mere 2\% shift in the integrated branching ratio at low-$q^2$~\cite{Huber:2020vup}.

Nonfactorizable diagrams can in principle be described by the $O(1/m_c^2)$ power corrections first discussed in Ref.~\cite{Buchalla:1997ky}. Unfortunately, as was discovered in an SCET-I analysis of inclusive modes at low-$q^2$, the virtual-photon emission from the charm loop introduces a dependence on the hard-collinear scale in the anti-collinear (with respect to $X_s$) direction. The resulting contributions to the SCET-I OPE appear in the subleading order and survive after integration over the low-$q^2$ region. These corrections are commonly referred to as {\it resolved photon contributions} and, while currently not known, have been estimated to be $O(5\%)$~\cite{Hurth:2017xzf, Benzke:2017woq, Benzke:2020htm}. 

\medskip

\noindent\underline{Inclusive decays at high-$q^2$} admit the same OPE as at low-$q^2$ with few crucial differences. First, the OPE is an expansion in $1/(m_b - \sqrt{q^2})$ and breaks down at the endpoint of the spectrum. This problem appears explicitly in the expressions for the power corrections which have integrable singularities at $q^2 = m_b^2$. Focusing on the high-$q^2$ integrated rate, the OPE breakdown results in large power corrections which completely dominate the theoretical uncertainty on the SM prediction. The solution to this problem is to consider the high-$q^2$ integrated branching ratio normalized to the $B\to X_u \ell\nu$ branching ratio measured with the same $q^2$ cut~\cite{Ligeti:2007sn}. The OPE breakdown in both numerator and denominator is almost identical and cancels in the ratio. The resulting observable has a power expansion which is under fairly good control~\cite{Huber:2020vup, Huber:2024rbw}. 

The second major difference with respect to the inclusive at low-$q^2$ is that for $q^2 > m_{\psi(2S)}^2$ we still need to integrate over four broad resonances (rather than having to deal   with the tail of the $J/\psi$ only). As at low-$q^2$, the factorizable contributions to these resonances are described using the KS method while nonfactorizable effects on the integrated rate are given by $O(1/m_c^2)$ which, in this case, are local and given in terms of standard HQET matrix elements~\cite{Buchalla:1997ky}. The presence of these broad resonances inside the integration region leads to fairly large effects: the replacement of the perturbative two loop charm contribution to the OPE with the KS ansatz yields a shift of about -10\% of the integrated branching ratio~\cite{Huber:2020vup}. The extent to which nonfactorizable effects in the production of the broad charmonium resonances above the $\psi(2S)$ are captured by local $O(1/m_c^2)$ corrections is currently under investigation.

\begin{figure}[t!]
    \centering
\includegraphics[width=7cm]{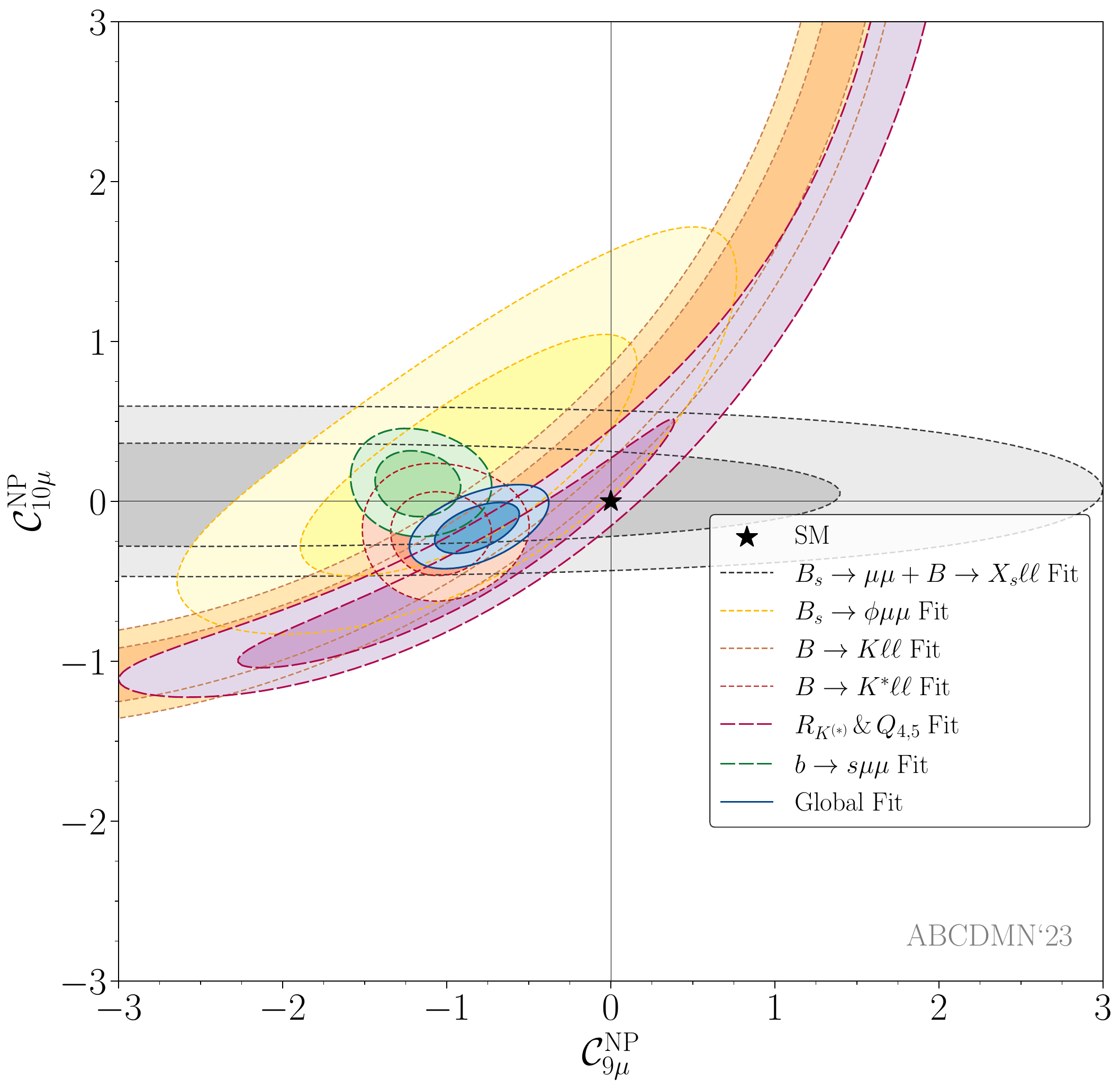}
\caption{Global fit taken from Ref.~\cite{Alguero:2023jeh}, 1$\sigma$ (dark-shaded) and 2$\sigma$ (light-shaded) confidence regions for $(\Cc{9\mu}^{\rm NP},\Cc{10\mu}^{\rm NP})$ scenarios. Distinct fits are performed separating each of the $b\to s\ell^+\ell^-$ by modes  (short-dashed contours), the LFUV observables and the  combined $b\to s\mu^+\mu^-$ modes (long-dashed contours), and the global fit (solid contours). 
}\label{fig:C9mu_C10mu}
\end{figure}

\section{Definition of observables and NP sensitivity} \label{sec:observables}

In this section we present two observables designed to discriminate between the two possibilities under scenario A: a universal NP contribution ${\cal C}_9^{\rm NP}$ or a hypothetical constant hadronic contribution ${\cal C}_{9, \rm extra}$. Notice, however, that for the analysis of each of the observables in this section we will not yet enforce universality, so we will keep each contribution ${\cal C}_{9, \rm extra}^{K^{(*)},\,\rm high\, or \, low}$ as formally independent.

The observables proposed in this paper are constructed from  combinations of isospin averaged branching ratios and are defined as follows:
\begin{align}
{\cal F}_{K^{(*)}}^{[1.1,6]}
&= \frac{{\cal B} (B\to K^{(*)} \mu\mu)_{[1.1,6]}}{{\cal B} (B\to X_s \ell\ell)_{[1,6]}}\; ,\label{F16}\\
{\cal F}_{K^{(*)}}^{>15}
&= \frac{{\cal B} (B\to K^{(*)} \mu\mu)_{>15}}{{\cal B} (B\to X_s \ell\ell)_{>15}} \; , \label{F15}
\end{align}
for $q^2\in [1.1,6]\; {\rm GeV}^2$ and $q^2 > 15\; {\rm GeV}^2$, respectively\footnote{Note that for the inclusive at low-$q^2$ we use the more conventional $[1,6]\; {\rm GeV}^2$ window for which all existing theoretical and experimental studies have been performed.}.

The key idea  behind the construction of these four quantities is that only exclusive modes are affected by hypothetically large non-local power corrections, which we indicate by ${\cal C}_{9,\rm extra}^{K^{(*)},\rm high\, or\, low}$, whereas both  exclusive and inclusive modes are affected by the universal NP contribution. 

For completeness in Fig.~\ref{fig:C9mu_C10mu} we reproduce the confidence regions in the $(\mathcal{C}_{9\mu}^\mathrm{NP},\mathcal{C}_{10\mu}^\mathrm{NP})$ plane from Ref.\cite{Alguero:2023jeh}. These show a  clear preference for a non-zero $\mathcal{C}_{9}^{\rm NP}$ contribution, and excellent compatibility between the global fit and inclusive $B \to X_s\ell\ell$ decays, despite the uncertainty in the determination from the inclusive modes being large.

We work under the assumption that the only uncontrolled theoretical uncertainty is the size of a hypothetical hadronic constant contribution to the exclusive channels which we parameterize as a constant shift to the Wilson coefficient of the $O_9$ operator: ${\cal C}_9 \to {\cal C}_9 + {\cal C}_{9, \rm extra}^{K^{(*)},\,\rm low\, or\, high}$. In this scenario  numerators and denominators of the observables in Eqs.~(\ref{F16}) and (\ref{F15}) are second order polynomials in ${\cal C}_9^{\rm NP}+ {\cal C}_{9, \rm extra}^{K^{(*)},\rm low\, or\, high}$ and ${\cal C}_9^{\rm NP}$, respectively.

It is clear that independent measurements of inclusive and exclusive modes have the ability to independently determine $\mathcal{C}_{9}^{{\rm NP}}$ and ${\cal C}_{9, \rm extra}^{K^{(*)},\, \rm low\, or\, high}$. This, for instance, can be seen from Fig.~5 of Ref.~\cite{Huber:2024rbw} where it is shown that inclusive modes can confirm the presence of a non-vanishing $\mathcal{C}_{9}^{\rm NP}$ (failure to observe a tension in inclusive modes would suggest a resolution of the exclusive anomalies in terms of ${\cal C}_{9, \rm extra}^{K^{(*)},\rm low\,or\,high} \neq 0$). See also the discussion in section~\ref{sec:F_exp}.

In this letter we argue that the observables ${\cal F}_{K^{(*)}}^{[1.1,6]}$ and ${\cal F}_{K^{(*)}}^{>15}$ have a reduced sensitivity to $\mathcal{C}_9^{\rm NP}$ given the partial cancellation between numerator and denominator  and this property can be used to identify the presence or not of a non-negligible ${\cal C}_{9, \rm extra}^{K^{(*)},\rm low\, or \, high}$ contribution to the exclusive mode. 

The ingredients we need for this analysis are the full error budgets and NP formulae for the various inclusive and exclusive branching ratios. 

The relevant formulae for the inclusive modes are~\cite{Huber:2020vup}:
\begin{align}
{\cal B} (B\to X_s \ell\ell)_{[1,6]} &= \left( 17.44 + 3.31\; \mathcal{C}_{9}^{\rm NP} + 0.56 \; {\mathcal{C}_{9}^{\rm NP}}^2 \right) \times 10^{-7} \; , \label{eq:BR16}\\
{\cal R} (B\to X_s \ell\ell)_{>15} &= \frac{{\cal B} (B\to X_s \ell\ell)_{>15}}{{\cal B} (B\to X_u \ell\nu)_{>15}} \; \nonumber \\
& =\left( 27.00 + 6.32 \; \mathcal{C}_{9}^{\rm NP} + 0.93 \; {\mathcal{C}_{9}^{\rm NP}}^2 \right) \times 10^{-4} \; . \label{eq:R15}
\end{align}
We assume that the uncertainties on these two observables are identical to those on the SM predictions: $\delta {\cal B} (B\to X_s \ell\ell)_{[1,6]} = 7.4\%$ and $\delta {\cal R} (B\to X_s \ell\ell)_{>15} = 7.2\%$. 

The low-$q^2$ branching ratio is the average of electron and muon channel, includes QED corrections and corresponds to what B-factories measure. 
The high-$q^2$ ratio, on the other hand, is optimized for LHCb and, thus, does not include any QED effect.  As we discussed in section~\ref{sec:excinc}, only the ratio ${\cal R}$ can be calculated reliably in terms of an OPE. The theoretical prediction for the high-$q^2$ branching ratio, required to construct the observables ${\cal F}_{K^{(*)}}^{>15}$, can be obtained by combining Eq.~(\ref{eq:R15}) with the experimental measurement of the semileptonic branching ratio at high-$q^2$, under the assumption of the absence of NP contributions to the latter. Using the $B\to X_u \ell\nu$ spectrum~\cite{Belle:2021ymg}, the high-$q^2$ $B\to X_u \ell\nu$ branching ratio is determined to be~\cite{Huber:2024rbw}:
\begin{align}
{\cal B} (B\to X_u \ell\nu)_{>15}^{\rm exp} = (1.52  \pm 0.28) \times 10^{-4} \; .
\end{align}
The high-$q^2$ branching ratio in the presence of NP contributions is, therefore:
\begin{align}
{\cal B} (B\to X_s \ell\ell)_{>15} 
& =\left( 4.10 + 0.96 \; {\cal C}_{9}^{\rm NP} + 0.14 \; {{\cal C}_{9}^{\rm NP}}^2 \right) \times 10^{-7} \; , \label{eq:BR15}
\end{align}
with a total uncertainty $\delta {\cal B} (B\to X_s \ell\ell)_{>15} = \sqrt{(7.2\%)^2 + (18.4\%)^2} = 19.8\%$.

The branching ratios of the exclusive modes can be expressed as quadratic functions of NP contribution to $\mathcal{C}_9^\mathrm{NP}$ and of a residual hadronic contribution, $\mathcal{C}_{9,\,\mathrm{extra}}^{K^{(*)},\rm low\, or\, high}$, which (if non-zero) is not accounted for by the calculation of the non-local form factor at LO  and our models for its NLO  contributions. In  scenario A we take these residual  hadronic contributions to be constant and equal to ${\cal C}_{9,{\rm extra}}$. Explicit expressions for the $K^{(*)}$ branching ratios at low- and high-$q^2$, following the approach in Ref.~\cite{Alguero:2023jeh}, are collected in Tables~\ref{table1} and \ref{table2}. 

The four observables ${\cal F}_{K^{(*)}}^{[1,6],>15}$  can now be easily constructed combining the expressions of the inclusive branching ratios given in Eqs.~(\ref{eq:BR16}) and (\ref{eq:BR15}) as well as the corresponding expressions for the exclusive modes given in Tables~\ref{table1} and \ref{table2}.
Note that the four ${\cal F}_{K^{(*)}}^{[1,6],>15}$ observables are given by ratios whose numerators  and denominators are quadratic polynomials in ${\cal C}_{9}^{{\rm NP}} + 
{\cal C}_{9, \rm extra}^{K^{(*)},\rm low\, or \, high}$ and ${\cal C}_{9}^{{\rm NP}}$, respectively. 

As we discuss below, these observables may offer a unique way to disentangle ${\cal C}_{9}^{{\rm NP}}$ from ${\cal C}_{9, \rm extra}^{K^{(*)}}$ at low- and high-$q^2$. Moreover, since they involve only ratios of partial widths, they can be measured entirely by LHCb without requiring external measurements of normalizing modes (e.g. $B\to J/\psi K^{(*)}$) from BaBar or Belle.

\begin{table}[th]
\centering
\begin{tabular}{lcccc}
\hline
& & & & \\[-2mm] 
& Region ($q^2$) & $\alpha$ & $\beta$ & $\gamma$\\[2mm]
\hline
& & & & \\[-2mm] 
$10^7 \times {\cal B} \left(B\to K\mu^+\mu^- \right)$ & $[1.1,6]$ & $1.727$ & $0.399$ & $0.051$\\[2mm]
$10^7 \times {\cal B} \left(B\to K\mu^+\mu^-\right)$ & $>15$ & $1.105$ & $0.252$ & $0.033$\\[2mm]
\hline
& & & & \\[-2mm] 
$10^7 \times {\cal B} \left(B\to K^*\mu^+\mu^-\right)$ & $[1.1,6]$ & $1.931$ & $0.342$ & $0.064$\\[2mm]
$10^7 \times {\cal B} \left(B\to K^*\mu^+\mu^-\right)$ & $>15$ & $2.640$ & $0.600$ & $0.082$\\[2mm]
\hline
\end{tabular}
\caption{Theoretical predictions for $10^7\times\mathcal{B}(B\to K^{(*)}\mu^+\mu^-)$ written in an approximate quadratic form $\alpha + \beta\,\tilde{\mathcal{C}}_9 + \gamma\,\tilde{\mathcal{C}}_9^2$, where $\tilde{\mathcal{C}}_9 = \mathcal{C}_9^{\rm NP}+
{\cal C}_{9, \rm extra}^{K^{(*)},\rm low\, or \, high}$
}\label{table1}
\end{table}
\begin{table}[th]
\centering
\begin{tabular}{lcccc}
\hline
& & & & \\[-2mm] 
& Region ($q^2$) & $\alpha$ & $\beta$ & $\gamma$\\[2mm]
\hline
& & & & \\[-2mm] 
$10^7 \times \delta {\cal B} \left(B\to K\mu^+\mu^- \right)$ & $[1.1,6]$ & $0.133$ & $0.032$ & $0.003$\\[2mm]
$10^7 \times \delta {\cal B} \left(B\to K\mu^+\mu^-\right)$ & $>15$ & $0.147$ & $0.034$ & $0.003$\\[2mm]
\hline
& & & & \\[-2mm] 
$10^7 \times \delta {\cal B} \left(B\to K^*\mu^+\mu^- \right)$ & $[1.1,6]$ & $1.036$ & $0.177$ & $0.035$\\[2mm]
$10^7 \times \delta {\cal B} \left(B\to K^*\mu^+\mu^- \right)$ & $>15$ & $0.235$ & $0.056$ & $0.005$\\[2mm]
\hline
\end{tabular}
\caption{Uncertainties $10^7\times \delta \mathcal{B}(B\to K^{(*)}\mu^+\mu^-)$ on the theoretical predictions for the various exclusive modes written in an approximate quadratic form $\alpha + \beta\,\tilde{\mathcal{C}}_9 + \gamma\,\tilde{\mathcal{C}}_9^2$, where $\tilde{\mathcal{C}}_9 = \mathcal{C}_9^{\rm NP}+
{\cal C}_{9, \rm extra}^{K^{(*)},\rm low\, or \, high}$.}\label{table2}
\end{table}

\section{Experimental determinations of \texorpdfstring{${\cal F}_{K^{(*)}}^{[1,6]}$}{Flow} and \texorpdfstring{${\cal F}_{K^{(*)}}^{>15}$}{Fhigh}}
\label{sec:F_exp}
In this section we show how to combine existing experimental measurements of various inclusive and exclusive modes to obtain the observables ${\cal F}_{K^{(*)}}^{[1,6]}$ and ${\cal F}_{K^{(*)}}^{>15}$. 

LHCb can only measure normalized ratios of the various exclusive $b\to s \mu\mu$ modes. For instance: 
\begin{align}
{\cal B} (B^+\to K^+ \mu^+\mu^-) = \left[\frac{\Gamma(B^+\to K^+ \mu^+\mu^-)}{\Gamma(B^+\to K^+J/\psi)} \right]_{\rm LHCb} \times \left[ {\cal B} (B^+ \to K^+ J/\psi)
\right]_{\rm external} \; .
\label{eq:bkmmstructure}
\end{align}
In order to propagate correctly uncertainties stemming from the charmonium normalizations, we first determine the experimental averages for the two factors in Eq.~(\ref{eq:bkmmstructure}) (similar decompositions hold for all the other modes) and then combine them. Averages for the normalizing charmonium modes are presented in App.~\ref{ssec:BRcharmonium}. The combinations of the LHCb and CMS measurements at low- and high-$q^2$ for the normalized ratios (i.e. the first factor in Eq.~(\ref{eq:bkmmstructure})), are presented in Apps.~\ref{ssec:normalizedbsmmlow} and \ref{ssec:normalizedbsmmhigh}, respectively. The averages of the low- and high-$q^2$ $b\to s\mu\mu$ branching ratios, obtained from the product of these two terms, are presented in Apps~\ref{ssec:bsmmlow} and \ref{ssec:bsmmhigh}.

For the construction of the various ${\cal F}$ ratios we need also averages for the inclusive $B\to X_s \ell^+\ell^-$ measurements. At low-$q^2$ we calculate the weighted average of the BaBar~\cite{BaBar:2004mjt, BaBar:2013qry}, Belle~\cite{Belle:2005fli, Belle:2014owz} and Belle~II~\cite{Belle2:Moriond2026} measurements of the inclusive branching ratio for $q^2 \in [1,6] \; {\rm GeV}^2$:
\begin{align}
    {\cal B} (B\to X_s \ell\ell)_{[1,6]} = (1.44 \pm 0.21) \times 10^{-6} \; ,
    \label{eq:BRXsexp}
\end{align}
where the electron and muon modes have been averaged. At high-$q^2$ we employ a completely different strategy which is based on the observation that the inclusive is essentially saturated by summing over the $K$, $K^*$, $(K\pi)_{\rm S-wave}$ and $K\pi\pi$ modes, all of which have been already measured at LHCb. The inclusive isospin-averaged high-$q^2$ branching ratio as sum of exclusive can be written as:
\begin{align} \label{expnumb}
    {\cal B} (B\to X_s \mu\mu)_{>15} &= {\cal B}(B\to K\mu\mu) +  {\cal B}(B\to K^*\mu\mu) + {\cal B}(B\to (K\pi)_S\mu\mu) + {\cal B}(B\to K\pi\pi\mu\mu) \nonumber \\
    &= (2.80 \pm 0.12 ) \times 10^{-7} ,
\end{align}
where the various contributions are given in Eqs.~(\ref{eq:BRK_high}), (\ref{eq:BRK*_high}), (\ref{eq:BRKpiS_high}) and (\ref{eq:BRKpipi_high}).

The ${\cal F}_{K^{(*)}}^{[1,6]}$ ratios can only be constructed by constructing separately the exclusive and inclusive branching ratios because the low-$q^2$ inclusive branching ratio is currently only accessible at B-factories. Using the isospin averages in Eqs.~(\ref{eq:BRK_low}) and (\ref{eq:BRK*_low}) we immediately obtain:
\begin{align}
{\cal F}_K^{[1.1,6]} &= 0.084 \pm 0.013 , \label{eq:FK_low} \\
{\cal F}_{K^*}^{[1.1,6]} &= 0.110 \pm 0.017. \label{eq:FK*_low}
\end{align}

At high-$q^2$, employing the decomposition of the inclusive branching ratio given in Eq.~(\ref{expnumb}), we can write:
\begin{align}
    {\cal F}_K^{>15} &=  \left[ 1 + \frac{
    {\cal B}(K^*\mu\mu) + {\cal B}((K\pi)_S \mu\mu) + {\cal B} (B \to K\pi\pi \mu\mu)
}{{\cal B}( K\mu\mu)} \right]^{-1} = 0.309 \pm 0.015, \label{eq:FK_high} \\
    {\cal F}_{K^*}^{>15} &=  \left[1+  \frac{
    {\cal B}(K\mu\mu) + {\cal B}( (K\pi)_S\mu\mu) + {\cal B} (B \to K\pi\pi \mu\mu)
    }{{\cal B}( K^*\mu\mu)} \right]^{-1} = 0.629 \pm 0.021, \label{eq:FK*_high}
\end{align}
where the various ratios of branching ratios are computed using the results in Eqs.~(\ref{eq:BRK_high}), (\ref{eq:BRK*_high}), (\ref{eq:BRKpiS_high}) and (\ref{eq:BRKpipi_high}), including the impact of correlations that stem from common normalizations.

\section{Current status and future projections}
\label{sec:current}
We are now ready to show how to use the four $\cal F$ observables that we introduced in Eqs.~(\ref{F16}) and (\ref{F15}) to discriminate between the interpretation of the anomalies in terms of ${\cal C}_9^{\rm NP}$ or hypothetical ${\cal C}^{K^{(*)},\rm low\, or\, high}_{9,\rm extra}$ contributions. Notice that the analysis that we will present here is done observable by observable and it allows to formally test the corresponding ${\cal C}^{K^{(*)},\rm low\, or\, high}_{9,\rm extra}$ associated with  each observable, so we keep this general notation in the plots.
%
\begin{table}[t]
\centering
\begin{tabular}{|c|c|c|c|c|} \hline
    & ${\cal F}_K^{>15}$ & ${\cal F}_{K^*}^{>15}$ & ${\cal F}_K^{[1.1,6]}$ & ${\cal F}_{K^*}^{[1.1,6]}$ $\vphantom{\Big(}$ \\ \hline
    $({\cal C}_9^{\rm NP}, {\cal C}_{9,\rm extra} )= (0,0)$     & 0.270(64) & 0.64(14)  & 0.099(11)  & {0.111(60)} $\vphantom{\Big(}$\\ \hline
    $({\cal C}_9^{\rm NP}, {\cal C}_{9,\rm extra} )= (-1.16,0)$ & 0.271(64) & 0.65(14)  & 0.0928(98)  & {0.113(62)} $\vphantom{\Big(}$\\ 
    $({\cal C}_9^{\rm NP}, {\cal C}_{9,\rm extra} )= (0,-1.16)$ & 0.210(50) & 0.50(11)  & 0.0764(81) & {0.093(51)} $\vphantom{\Big(}$\\ \hline
    experiment                                                  & 0.309(15) & 0.629(21) & 0.084(13)  & 0.110(17) $\vphantom{\Big(}$\\ \hline
    \end{tabular}
    \color{black}
    \caption{Theoretical determinations of the ratios ${\cal F}_{K^{(*)}}^{[1.1,6],>15}$ for various values of ${\cal C}_9^{\rm NP}$ and ${\cal C}_{9,\rm extra}$ and comparison to their current experimental determinations. Notice that, by construction, the sensitivity to NP is almost flat.}
    \label{tab:F_th_exp}
\end{table}
However, the preference of global fits, as well as of the bin-by-bin and mode-by-mode analysis done in Ref.~\cite{Alguero:2023jeh} (see also Ref.~\cite{Bordone:2024hui}) for a universal constant contribution coming from either NP or from a hypothetical constant hadronic contribution, imposes universality. Therefore, the results of the global fits can be interpreted as a non vanishing ${\cal C}_9^{\rm NP}$ or a universal ${\cal C}_{9,\rm extra}$ equal to $-1.16 \pm 0.17$ \cite{Alguero:2023jeh}. In order to discriminate between the two possibilities we compare the measured value of the four ${\cal F}$ ratios, given in Eqs.~(\ref{eq:FK_low}), (\ref{eq:FK*_low}), (\ref{eq:FK_high}) and (\ref{eq:FK*_high}), to the theoretical predictions obtained from the formulae given in Sec.~\ref{sec:observables} under the assumptions $({\cal C}_9^{\rm NP}, {\cal C}_{9,\rm extra})$ = $(-1.16,0)$ or $(0,-1.16)$. The results are summarized in Table~\ref{tab:F_th_exp}. In particular, the two high-$q^2$ ratios show a slight preference for ${\cal C}_9^{\rm NP}$ at $1.9\sigma$ ($1.2\sigma$) for the $K$ ($K^*$) final state; the two low-$q^2$ ratios are compatible with both scenarios at less than $1\sigma$. Note that the uncertainties on the theoretical predictions for different values of ${\cal C}_9^{\rm NP}$ and ${\cal C}_{9,\rm extra}$ are 100\% correlated. 

\begin{figure}[t]
\begin{center}
\includegraphics[width = 0.48\textwidth]{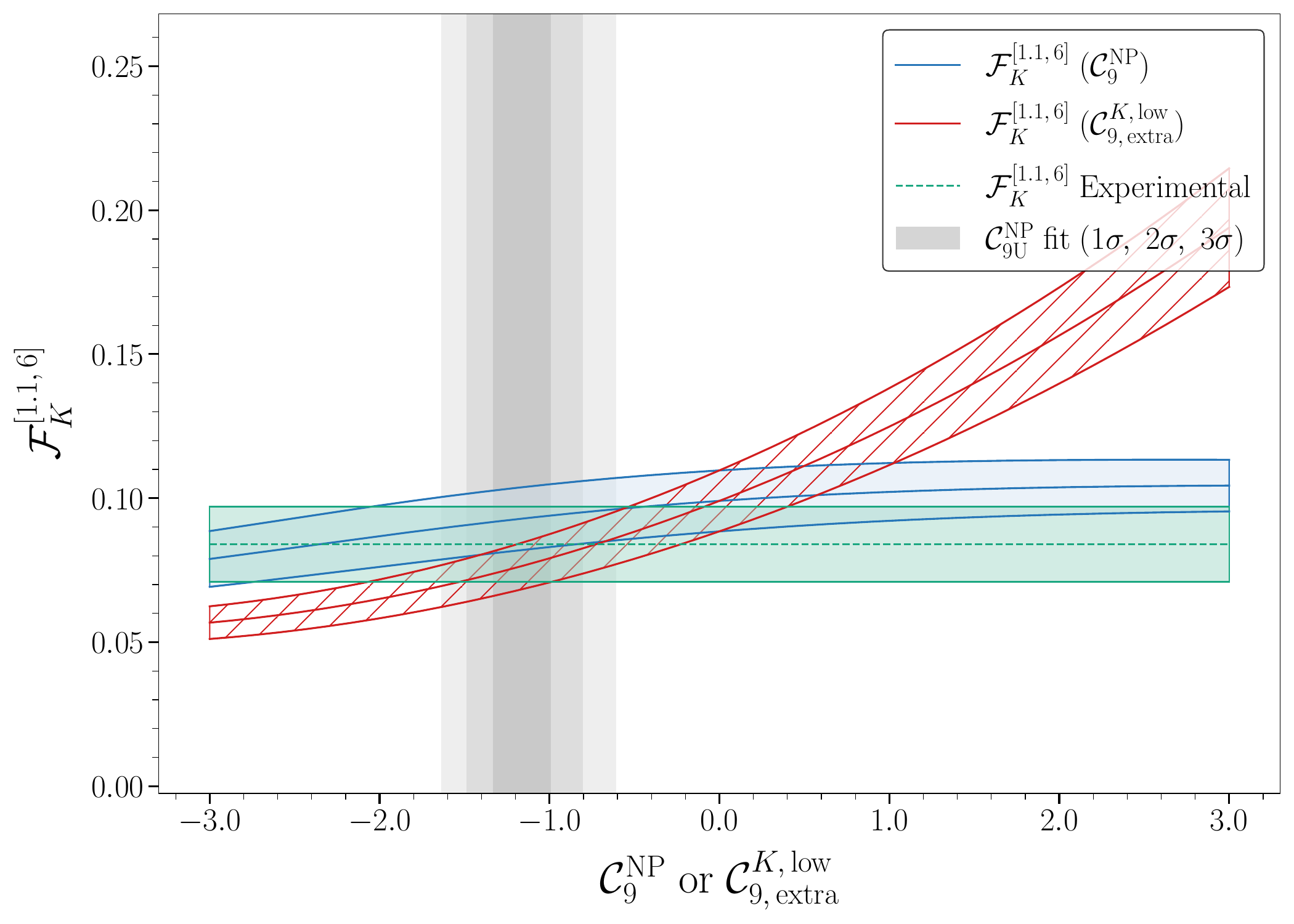}
\quad 
\includegraphics[width = 0.48\textwidth]{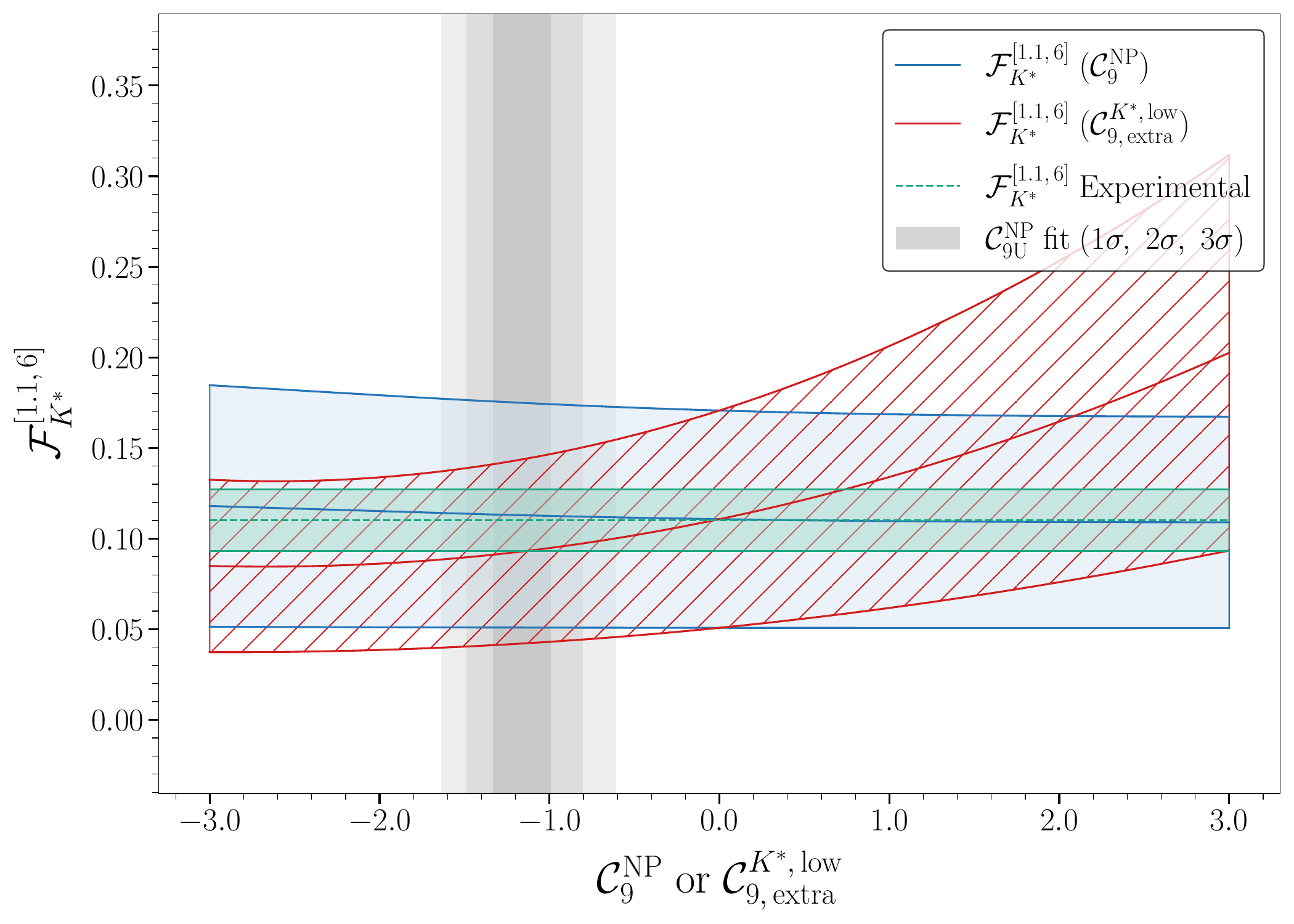} \\[2mm]
\includegraphics[width = 0.48\textwidth]{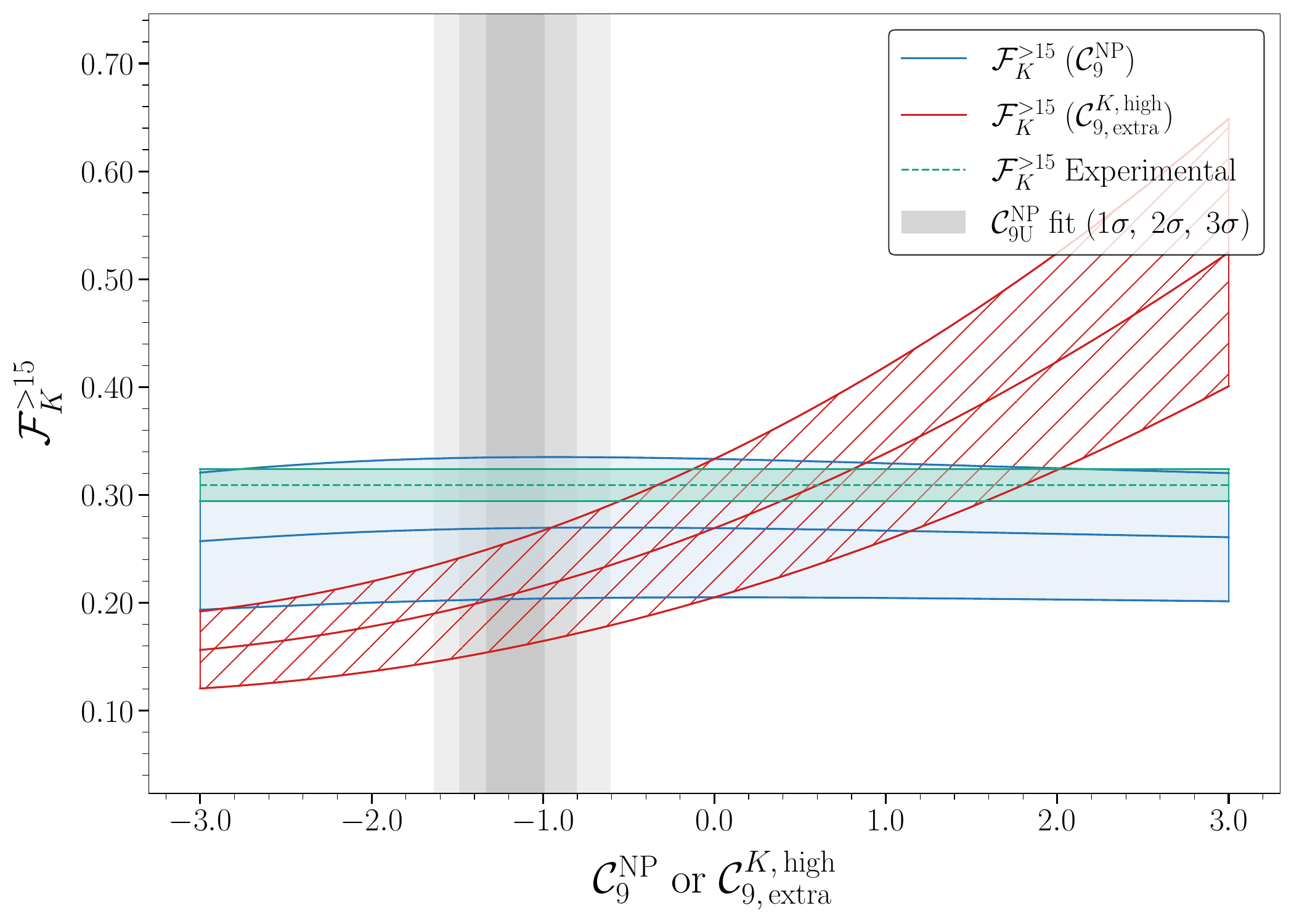}
\quad 
\includegraphics[width = 0.48\textwidth]{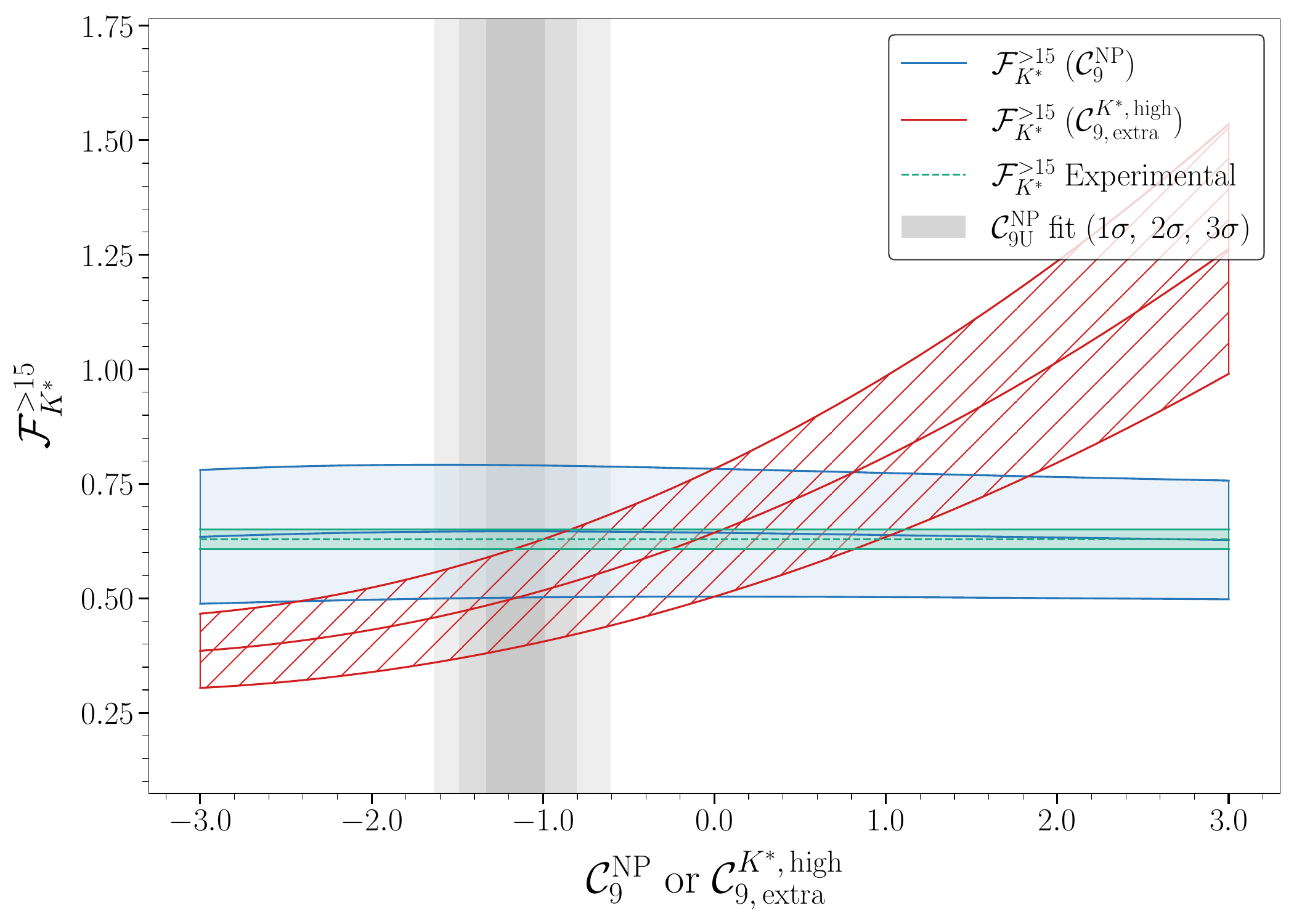}
\caption{Current status of experimental and theoretical determinations of the observables ${\cal F}_{K^{(*)}}^{[1.1,6],>15}$. The horizontal dashed green lines correspond to the current experimental determinations (and the corresponding 68\%CL allowed range) given in Eqs.~(\ref{eq:FK_low}), (\ref{eq:FK*_low}), (\ref{eq:FK*_high}) and (\ref{eq:FK*_high}). The vertical gray bands indicate the 1$\sigma$, 2$\sigma$ and 3$\sigma$ results of the global fits to exclusive data at low- and high-$q^2$ that can be interpreted in terms of ${\cal C}_9^{\rm NP}$ or ${\cal C}^{K^{(*)},\rm low\,or\, high}_{9,\rm extra}$. The red solid lines (and the corresponding 68\%CL allowed range) are the theoretical predictions ${\cal F}^{[1.1,6],>15}_{K^{(*)}} ({\cal C}_9^{\rm NP}=0, {\cal C}_{9,\rm extra}^{K^{(*)},\rm low\,or\,high} )$. The blue solid line (and the corresponding 68\%CL allowed range) is the theoretical prediction ${\cal F}^{[1.1,6],>15}_{K^{(*)}} ({\cal C}_9^{\rm NP}, {\cal C}_{9,\rm extra}^{K^{(*)},\rm low\,or\,high}=0)$. 
}
\label{fig:Fcomparison}
\end{center}
\end{figure}

In Fig.~\ref{fig:Fcomparison} we decouple the analysis presented in Tab.~\ref{tab:F_th_exp} from the actual results of the global fits. We consider all four $\cal F$ ratios and for each we show the current experimental determination (dashed green lines) and the corresponding theoretical predictions for the ${\cal C}_9^{\rm NP}\neq 0$ and ${\cal C}_{9,\rm extra}\neq 0$ scenarios (blue and red lines, respectively); the green, blue and hashed bands correspond to the allowed 68\% confidence level (CL) ranges. In each plot the gray bands display the 1$\sigma$, 2$\sigma$ and 3$\sigma$ ranges allowed by the global fit. 

The discriminating power of each $\cal F$ ratio depends crucially on the uncertainties on the two theoretical predictions (for ${\cal C}_9^{\rm NP} \neq 0$ and ${\cal C}_{9,\rm extra} \neq 0$) and on its experimental determination. In the balance of this section we discuss future prospects based on the current recorded luminosity at LHCb and Belle II and their future expectations. 

Let us begin with a discussion of the experimental measurements of these ratios. 

At low-$q^2$ the dominant uncertainty on ${\cal F}_K^{[1.1,6]}$ and ${\cal F}_{K^*}^{[1.1,6]}$ stems from the inclusive $B\to X_s \ell\ell$ measurement which has been recently presented by Belle II~\cite{Belle2:Moriond2026} using $400\; {\rm fb}^{-1}$ of integrated luminosity. This measurement is currently statistically limited and, with about $1\; {\rm ab}^{-1}$ (which is expected by the end of the year) statistical and systematic components of the uncertainty will be about identical. The latter are dominated by the current determinations of the $B\to K^{(*)}\ell\ell$ branching ratios and by modeling of the hadronic $X_s$ system~\cite{Belle2:Moriond2026}. A conservative estimate based on $1\; {\rm ab}^{-1}$ integrated luminosity yields an 8.5\% projected uncertainty on the world average. This corresponds to about 10\% expected uncertainties on both ${\cal F}_K^{[1.1,6]}$ and ${\cal F}_{K^*}^{[1.1,6]}$ in the very near future. 

At high-$q^2$ the situation is much brighter because the experimental uncertainties can be improved with LHCb data only. In a first scenario, the ${\cal F}$ ratios are measured following the procedure described in Sec.~\ref{sec:F_exp}; i.e. with improved determinations of the various {\it normalized} exclusive rates but without any additional measurement of the normalizing charmonium modes. A projection based on a scenario in which the  uncertainties on the normalizing charmonium modes are kept at the current level and dominate the exclusive high-$q^2$ branching ratios that enter the $\cal F$ ratios, yields a reduction of the overall uncertainties from 4.9\% to 3.1\% and from 3.3\% to 1.4\% for ${\cal F}_K^{>15}$ and ${\cal F}_{K^*}^{>15}$, respectively. 

In a second scenario, the $\cal F$ ratios are constructed directly by LHCb by combining ratios of the various $b\to s\mu\mu$ exclusive modes. A complication arises because the poor knowledge of the $b$-quark fragmentation fractions allows only precise measurements of same charge modes, e.g. $\Gamma(B^+ \to K^{*+}\mu\mu)/\Gamma(B^+ \to K^{+}\mu\mu)$ and $\Gamma(B^0 \to K^{*0}\mu\mu)/\Gamma(B^0 \to K^{0}\mu\mu)$. There is, therefore, a trade-off between the reduced uncertainty due to the missing external normalizations and the need to consider modes with neutral pseudoscalar mesons in the final state. Focusing, for instance, on the $K^*/K$ isospin averaged ratio the two scenarios correspond to:
\begin{align}
\dfrac{{\cal B}(B\to K^* \mu\mu)}{{\cal B}(B\to K \mu\mu)}
=
\begin{cases}
\dfrac{
\langle 
{\cal B} (K^{*0}\mu\mu),{\cal B} (K^{*+}\mu\mu)
\rangle
}{
\langle 
{\cal B} (K^{0}\mu\mu),{\cal B} ( K^{+}\mu\mu)
\rangle
}
\simeq 
\dfrac{{\cal B} (K^{*0}\mu\mu)}{{\cal B} (K^{+}\mu\mu)}
=
\dfrac{{\cal R} ( K^{*0})}{{\cal R} (K^{+})}
\dfrac{{\cal B} (K^{*0}J/\psi)}{{\cal B} (K^{+}J/\psi)},\\\\
\left\langle 
\dfrac{{\cal B} (K^{*0}\mu\mu)}{{\cal B} (K^{0}\mu\mu)}
,
\dfrac{{\cal B} (K^{*+}\mu\mu)}{{\cal B} (K^{+}\mu\mu)}
\right\rangle,
\end{cases}
\end{align}
where the two ratios in the second scenario do not need any external normalization. In the experimental determination of these two ratios many sources of systematics are expected to cancel; thus an estimate of the projected experimental uncertainties requires a dedicated analysis.

Let us now comment on future prospects for the uncertainties on the theoretical determinations of the ${\cal F}$ ratios. 

The numerators of these ratios are given by the exclusive $B\to K^{(*)} \mu\mu$ branching ratios: the only major improvement expected in the near future will come from the calculation of the $B\to K^*$ form factors in lattice-QCD calculations that take into account the resonant nature of the $K^*$.\footnote{A first calculation of the form factors into a light vector meson ($B\to \rho$) has been presented in Ref.~\cite{Leskovec:2025gsw} (albeit for a heavy pion, $m_\pi = 320$ MeV). The resonance parameters of the $K^*$ have been calculated at the physical point in Refs.~\cite{Boyle:2024hvv, Li:2026qzg}. Currently, two groups (Jevsenak, Meinel, Leskovec; Black, Boyle, Di Carlo, Erben, Gülpers, Hansen, Lachini, Mukherjee, Portelli, Tsang) are working towards the calculation of the $B\to K^*$ form factors at the physical point.} 

The denominators of these ratios are given by the inclusive $B\to X_s \ell\ell$ branching ratios at low- and high-$q^2$. While the theoretical uncertainty on the branching ratio at low-$q^2$ is not expected to improve, the high-$q^2$ result will benefit from three near future developments. As discussed in Secs.~\ref{subsec:inclusive} and \ref{sec:observables}, we construct the high-$q^2$ branching ratio as the product ${\cal R}(B\to X_s \ell\ell)_{>15} \times {\cal B} (B\to X_u\ell\nu)_{>15}^{\rm exp}$. First, Belle~II is expected to present an improved measurement of the $B\to X_u \ell\nu$ branching ratio for $q^2 > 15 \; {\rm GeV}$; in a recent analysis~\cite{Belle-II:2025pye}, Belle~II measured the integrated branching ratio for $q^2 > 8 \; {\rm GeV}^2$ with an uncertainty that is roughly a factor of two better than the corresponding Belle measurement we use. Second, the largest parametric uncertainty in the calculation of the ratio ${\cal R}$ is the CKM matrix element $|V_{ub}|$, whose extraction from exclusive and inclusive $b\to u \ell\nu$ transitions is also expected to improve in the near future, thanks to new lattice calculations of the $B\to \pi$ form factors and updates to the inclusive determinations at NNLO accuracy~\cite{Broggio:2026edk,Chen:2026gin}. Finally, the single largest uncertainty on $\cal R$ stems from our poor knowledge of certain vacuum annihilation matrix elements which have been recently calculated, in the $D_s$ sector, using lattice-QCD~\cite{Black:2026rbz, Black:2026dzp}; the corresponding matrix elements for $B$ mesons are more complicated but possible to calculate within the same framework.

In conclusion, it is very likely that the uncertainties on the theoretical determinations of the high-$q^2$ ${\cal F}$ ratios will improve by more than 50\% in the near future. 

\section{Comparison with data of inclusive and exclusive at high-\texorpdfstring{$q^2$}{q2}}\label{sec:inclusive-exclusive}

At this point, we should remind the reader that one of the main questions we want to address in this paper is: can all observed deviations be explained solely by an unknown universal hadronic effect that mimics NP, or is genuine NP required and unavoidable? Obviously, even if a small non‑zero NP contribution is needed to explain all data or part of it, it would imply that a hadronic explanation alone is insufficient, and NP would necessarily be present. Therefore the answer to the question will be in that case {\it no} and the presence of NP will be unavoidable.
These observations together with the preference of global fits mentioned in the previous section establish our  main starting point: we assume here the presence  of a {\it constant universal} hadronic contribution ${\cal C}_{9,\rm extra}$ that mimics NP as an explanation of all the anomalies and explore its consequences.

Our complementary strategy here to disentangle a hypothetical hadronic contribution from a potential NP effect is to compare the theoretical prediction for the inclusive rate with the sum of the exclusive decay channels, and confront both with experimental data. This approach should be viewed as an additional, somewhat more rudimentary test than the ${\cal F}$-observables introduced in the previous sections, less precise, perhaps, but still able to provide interesting  indications.

In this section, we restrict our attention to the high‑$q^2$ region (above 15 GeV$^2$), which has the key property that the sum of the branching ratios ${\cal B}({B \to K^*\ell\ell})$ and ${\cal B}({B\to K\ell\ell})$ essentially saturates the inclusive SM prediction. In addition, there is a small residual contribution from the remaining channels, estimated in Ref.~\cite{Isidori:2023unk}. Using the same inputs (Wilson coefficients, quark masses, CKM parameters, normalizations, etc.) employed for the other exclusive channels, we obtain for the dependence of this residual branching ratio on $C_9^{\rm NP}$ and on the universal ${\cal C}_{9, \rm extra}$
\begin{equation} \label{eq:brest}
10^7\times {\cal B}_{\rm rest>15}
= 0.547
+0.124\,  ({\cal C}_9^{\rm NP}+{\cal C}_{9, \rm extra})
+0.015\, ({\cal C}_9^{\rm NP}+{\cal C}_{9, \rm extra}
)^2\,.
\end{equation}
This contribution that amounts to 13\% of the total sum of exclusive channels is more uncertain than the other exclusive channels, but we find very remarkable and non-trivial that it provides the precise amount needed, as discussed below, to reproduce almost exactly the $\mathcal{C}_9$-dependence of the inclusive mode. One can also easily see, by inspection of the relative weights of the coefficients (linear versus quadratic) in Eq.~\eqref{eq:brest}, that they are similar to the relative weights appearing in the other exclusive channels shown in Tables~\ref{table1} and~\ref{table2}.

\begin{table}[th]
    \centering
    \begin{tabular}{lcccc}\hline  
        & & & & \\[-2mm] 
        & Region ($q^2$)  & $\alpha$ & $\beta$ & $\gamma$ \\[2mm] \hline
        & & & & \\[-2mm] 
         $10^7 \times {\cal B}\left(B\to K \mu\mu\right)$              & $>15$ & 
          1.105 & 0.252 & 0.033 \\
         $10^7 \times {\cal B}\left(B\to K^* \mu\mu\right)$            & $>15$ & 
         2.640 & 0.600 & 0.082        \\ 
         $10^7 \times {\cal B}\left(B\to K\pi \mu\mu\right)$           & $>15$ & 0.547 & 0.124 & 0.015 \\[2mm] \hline
        & & & & \\[-2mm] 
         $10^7 \times {\cal B}\left(B\to X_s \mu\mu\right)$ (sum excl) & $>15$ & 4.292 & 0.976 & 0.130 \\ 
        & & & & \\[-2mm] \hline\hline
        & & & & \\[-2mm] 
         $10^7 \times {\cal B}\left(B\to X_s \mu\mu\right)$ (incl)     & $>15$ & 4.10  & 0.96  & 0.14  \\[2mm] \hline
    \end{tabular}
    \caption{Theoretical predictions for the inclusive $B\to X_s \mu^+\mu^-$ branching ratios in units of $10^{-7}$, obtained summing over the $K$, $K^*$ and $K\pi$ modes (sum excl) and within an OPE framework (incl). Each branching ratio is written as a quadratic form $\alpha + \beta \; \tilde {\cal C}_9 + \gamma \; \tilde {\cal C}_9^2$ where $\tilde{\mathcal{C}}_9 = \mathcal{C}_9^{\rm NP}+{\cal C}_{9, \rm extra}$ (assuming universality) for the sum of exclusive and $\tilde{\mathcal{C}}_9 = \mathcal{C}_9^{\rm NP}$ for the inclusive. In the first three rows we show the breakdown of the sum of exclusive in its three components. 
    }
    \label{tab:incl_vs_sumexcl}
\end{table}

The first step is the comparison between the theoretical expression for the inclusive branching ratio, given in Eq.~\eqref{eq:BR15} as a function of ${\cal C}_9^{\rm NP}$ and evaluated following subsection~\ref{subsec:inclusive}, and the prediction obtained by summing the exclusive decays, evaluated according to subsection~\ref{subsec:exclusive}. For the latter, we find
\begin{align} \label{eq:inclSM}
10^7 \times {\cal B}^{\sum\mathrm{exc}} = 10^7 \times \sum_{\mathrm{mode}}\mathcal{B}_{\mathrm{mode}>15} &=10^7 \times \left({\cal B}_{B\to K\ell\ell>15}+{\cal B}_{B\to K^{*0}\ell\ell>15}+{\cal B}_{rest>15}\right) \nonumber \\
& \simeq 4.29+0.98\, ({\cal C}_9^{\rm NP}+{\cal C}_{9, \rm extra}
)+0.13\, ({\cal C}_9^{\rm NP}+
{\cal C}_{9, \rm extra}
)^2,
\end{align}
with an uncertainty of $\delta=\pm 0.39$. These results are neatly summarized in Tab.~\ref{tab:incl_vs_sumexcl} where we show the comparison, as a function of ${\cal C}_9$, between the inclusive branching ratio $B\to X_s \mu^+\mu^-$ at high-$q^2$ calculated as a sum of exclusive modes ($K$, $K^*$ and $K\pi$) and using the OPE approach. Note the remarkable agreement between the two calculations and how the three exclusive contributions combine, for each power of ${\cal C}_9$, to reproduce the result of the OPE calculation. 

The comparison between the inclusive and the sum of exclusive will be done at two different levels, comparing, first, directly the branching ratios and second, comparing the ${\cal C}_9$-dependence of the inclusive versus the exclusive that has a smaller sensitivity on input parameters.

\begin{figure}[!th]
\begin{center}
\includegraphics[width=0.65\textwidth]{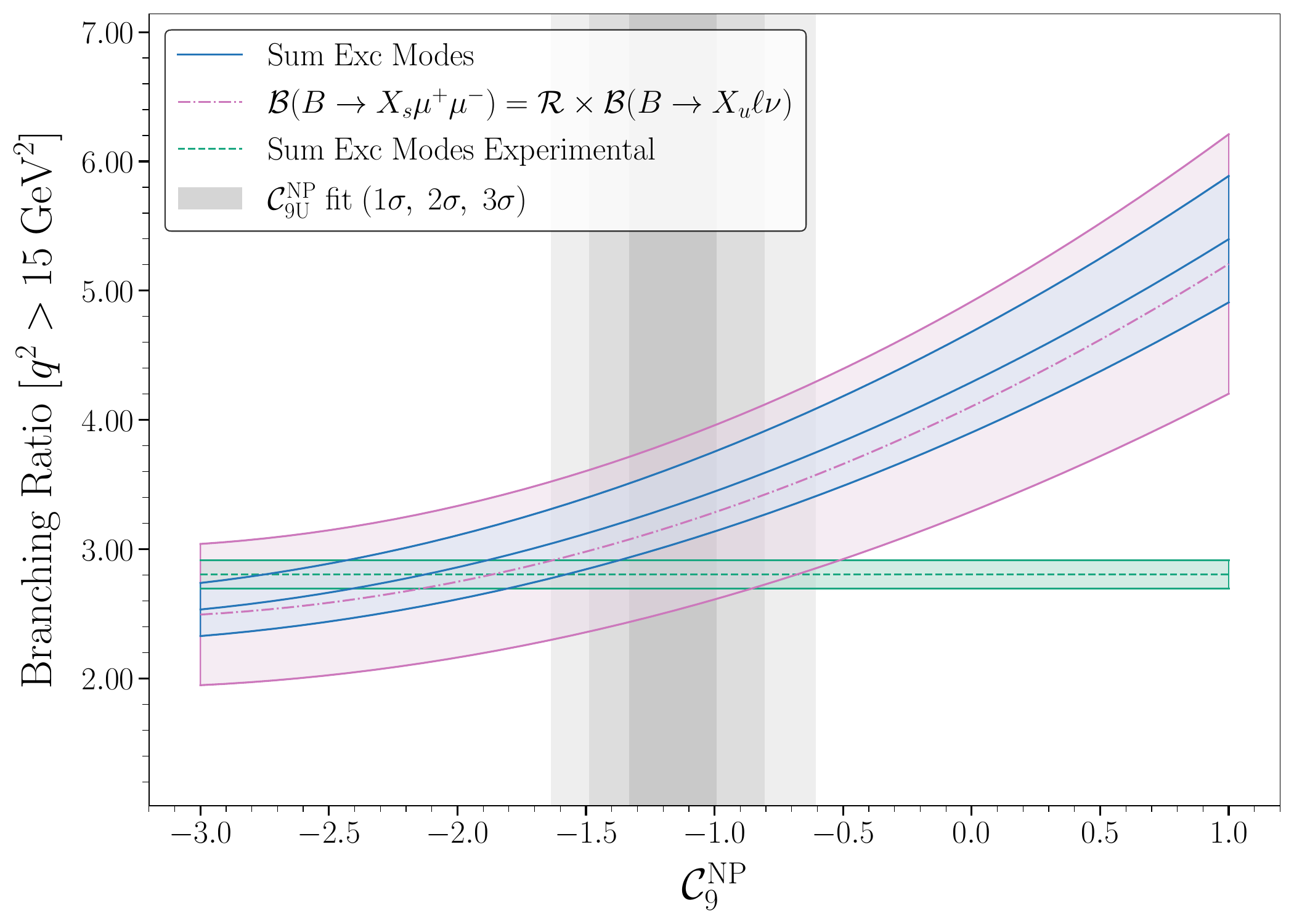}
\end{center}
 \caption{$B\to X_s \mu\mu$ branching ratio for $q^2>15\; {\rm GeV}^2$ calculated using the OPE (${\cal R} \times {\cal B}_{u\ell\nu}$) or as a sum of exclusive and comparison with the experimental determination.}
\label{fig:plot}
\end{figure}

Let's start with the branching ratios.
The main point  is that the same universal parameter ${\cal C}_{9, \rm extra}$ should fulfill two different conditions: 

\begin{itemize}

\item On the one side, ${\cal C}_{9,\rm extra}$ has to be in agreement with experimental data
\begin{equation}
\label{eq:eq2}
({\cal B}^{\sum \rm exc}_{\rm SM}-{\cal B}^{\sum \rm exc, exp})=0,
\end{equation}
where in the absence of ${\cal C}_{9, \rm extra}$ and ${\cal C}_{9}^{\rm NP}$ the sum of exclusive exhibits a tension of 3.7$\sigma$ between data ${\cal B}^{\sum \rm exc, exp}$ given by Eq.~\eqref{expnumb} and SM. Therefore this would require in the absence of NP that ${\cal C}_{9,\rm extra}=-1.16\pm 0.17$. For the inclusive mode, the deviation from the data amounts to 1.6$\sigma$, owing to its larger uncertainty. 

\item On the other side, ${\cal C}_{9,\rm extra}$ has to be adjusted to provide agreement between the inclusive and the sum of exclusive computation at all points (with or without NP) but particularly, for the SM prediction (${\cal C}_9^{\rm NP}=0$)
\begin{equation} \label{eq:eq1}
 ({\cal B}^{\sum \rm exc}_{\rm SM} -
{\cal B}^{\rm inc}_{\rm SM})=
0.
\end{equation}
Notice that Eq.~\eqref{eq:BR15} and Eq.~\eqref{eq:inclSM} rely on very different computational frameworks, yet their predictions agree remarkably well.  Fig.~\ref{fig:plot} shows excellent agreement between the inclusive prediction and the sum of exclusive modes in the presence of NP. If we impose that both computations must agree one finds at 68\% CL 
\begin{equation}\label{eq:c9extraie}
{\cal C}_{9,\rm extra}=-0.20^{+0.90}_{-1.10}.
\end{equation}
This theoretical comparison shows a preference for a small absolute value, compatible with zero, compared with the global fit preferred value of ${\cal C}_9^{\rm NP}\sim -1.16\pm 0.17$ although the uncertainty is still large. Moreover, the profile is highly non-parabolic and the likelihood very asymmetric, i.e, going beyond the 68\% CL the range spans up to very large values. Also there is a second solution which again implies a very large negative value ($\sim -7$) that would be in stark contradiction with the preferred values of ${\cal C}_9^{\rm NP}\sim -1.16\pm 0.17$ to explain all the anomalies. Interestingly, the uncertainty in Eq.~\eqref{eq:c9extraie} is mostly driven by the uncertainty in the inclusive computation. If we assume a reduction of 50\% in this error as discussed in the previous section and the central value remains untouched one finds at 68\% CL
\begin{equation}
{\cal C}_{9,\rm extra}=-0.20^{+0.59}_{-0.62}.
\end{equation}
In this case, the tails of the profile likelihood are more strongly constrained, and at the 95\% CL one obtains
${\cal C}_{9,\rm extra}=-0.20^{+1.18}_{-1.30}$. Therefore it is very important to improve on the inclusive uncertainty by improving the measurement of ${\cal B}(B\to X_u\ell\nu)$.

\end{itemize}

\noindent To summarize the comparison at the level of branching ratios: the universality pointed by data  disfavors the residual extra ${\cal C}_{9, \rm extra}$ contribution as a solution of all $B$-flavour anomalies, once we impose the condition that the sum of exclusive branching ratios should be equal to  the inclusive branching ratio. Therefore, one would need to break this universality, contrary to what the data seem to indicate.

\begin{figure}[!th]
\begin{center}
\includegraphics[width=0.65\textwidth]{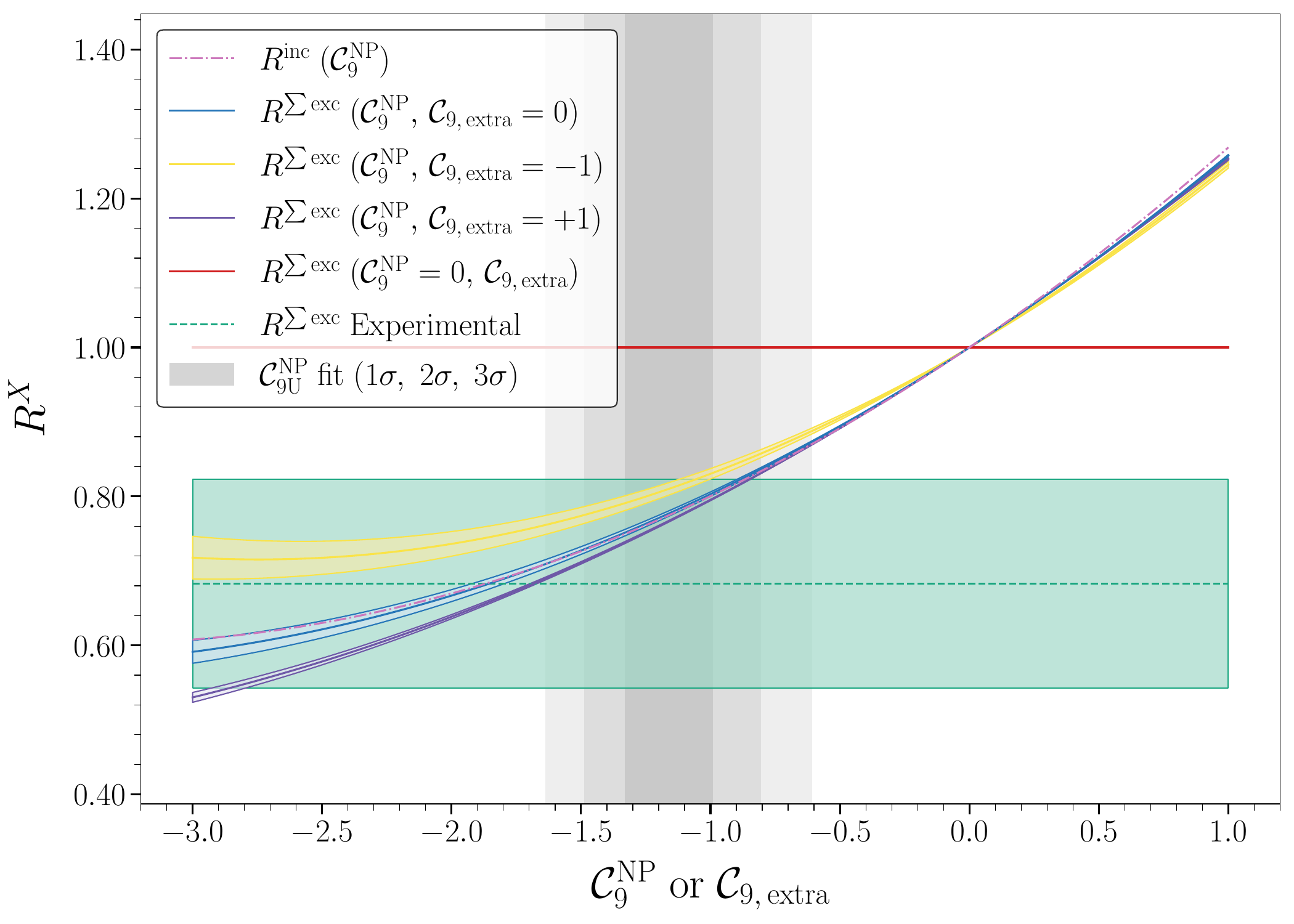}
\end{center}
\caption{
A more detailed comparison of the ${\cal C}_9$-dependence of the inclusive and exclusive modes under different hypotheses. Green corresponds to the experimental result, $R^{\rm exp}$, the red line corresponds to the absence of NP with only hadronic contributions, i.e, $R^{\sum \rm exc}({\cal C}_{9}^{\rm NP}=0, {\cal C}_{9,\rm extra})$, and the blue, yellow and magenta curves correspond to ${R}^{\sum exc}({\cal C}_9^{\rm NP},{\cal C}_{9, \rm extra}=0)$ and ${R}^{\sum exc}({\cal C}_9^{\rm NP},{\cal C}_{9, \rm extra}=\mp 1)$ respectively. This shows that when ${\cal C}_{9,\rm extra}$ is switched on (magenta and yellow bands) the necessary good agreement between inclusive and sum of exclusive breaks down.}
\label{fig:plotbis}
\end{figure}

We now turn to the comparison of the $\mathcal{C}_9$ dependencies of the inclusive mode and the sum of exclusive modes. Part of the remaining difference between the inclusive and the sum of exclusive modes arises from small differences in the input parameters used in the two evaluations. To perform a different comparison at the level of ${\cal C}_{9}$-dependencies in order to reduce normalization‑dependent effects, it is convenient to define the following ratios:
\begin{eqnarray} \label{eq:expressions}
R^{\rm inc}&=&\frac{{\cal B}^{\rm inc}}{{\cal B}^{\rm inc}_{\rm SM}}=1+0.234\, {\cal C}_9^{\rm NP}+0.034 \, {{\cal C}_9^{\rm NP}}^2,\nonumber \\
R^{\sum {exc}}&=& \frac{{\cal B}^{\rm \sum exc}}{{\cal B}^{\rm \sum exc}_{\rm SM}}=\frac{1+0.227 ({\cal C}_9^{\rm NP}+
{\cal C}_{9, \rm extra})+0.030 ( {\cal C}_9^{\rm NP}+{\cal C}_{9, \rm extra})^2}{1+0.227 {\cal C}_{9, \rm extra}+0.030 {\cal C}_{9, \rm extra}^2}.
\end{eqnarray}
The two limiting scenarios of interest are: i) ${\cal C}_9^{\rm NP}=0$, the hypothetical hadronic contribution entirely mimics NP, implying $R^{\sum {exc}}=1$, and ii) ${\cal C}_{9, \rm extra} = 0$, the purely NP case. Here we can compare the two expressions in Eq.~\eqref{eq:expressions}, for which the coefficient of the dominant linear term differs by only 2.8\% between the inclusive and exclusive determinations.

The corresponding experimental value is
\begin{equation}
R^{\rm exp}
= \frac{{\cal B}^{\rm \sum exc,exp}}{{\cal B}^{\rm inc}_{\rm SM}}
= 0.68 \pm 0.14,
\end{equation}
where the uncertainty includes $\delta {\cal B}^{\rm inc}_{\rm SM}$ from Eq.~(\ref{eq:BR15}).

This is illustrated in Fig.~\ref{fig:plotbis}. The two curves corresponding to $R^{\rm inc}$ and $R^{\rm \sum exc}$ in the NP‑only scenario overlap almost perfectly, confirming the consistency between the predictions for the inclusive decay and sum of exclusive channels. Both curves are in perfect agreement with the NP global fit solution (grey bands) of Fig.~\ref{fig:plotbis}. Instead, the  purely hadronic contribution, i.e,  $R^{\sum \rm exc}=1$ is in tension with the experimental band at { 2.3$\sigma$}. One also can observe in Fig.~\ref{fig:plotbis} that when ${\cal C}_{9,\rm extra}$ is switched on the tension between the $R^{\rm inc}$ and $R^{\sum exc}$ increases for large negative values of NP if ${\cal C}_{9,\rm extra}\sim -1$ (yellow band). The tiny uncertainty band of the $R$ ratios is a direct consequence of the strong correlation between the theory predictions for the numerator and denominator.

In summary, this alternative approach shows: i) on the one side, again a preference for NP in front of a constant hadronic universal contribution to explain the data in agreement with what the first analysis of the  ${\cal F}$ observables (still with large uncertainties) seems to point to, and ii) on the other side, that a universal unknown hadronic contribution has a different signature from a pure NP contribution, when the inclusive mode is included.

\section{Conclusions and Outlook} 
\label{sec:conclusions}

Since the observation of the first  $B$-flavour anomalies in semileptonic decays it has become customary to question whether these deviations arise from genuine Physics beyond the SM or from unaccounted hadronic effects that reflect our limited understanding of QCD. On the experimental side, the anomalies found in the $b\to s \mu\mu$ mediated decays have been experimentally confirmed in a persistent and coherent manner, not so in the case of Lepton Flavour Universality Violating observables. Moreover, global fits to a large set of semileptonic observables exhibit a clear preference for a single universal NP contribution to the Wilson coefficient of the operator $O_9$. 

In this analysis our aim has been to discern whether such a NP contribution is unavoidable to explain all the anomalies.
Consequently, if a hadronic contribution is able to mimic a universal NP shift in ${\cal C}_9$, it would have to do so consistently across all kinematic regions: in every $q^2$ bin, both at low and high recoil, and for all decay channels sensitive to this coefficient. This requirement leads to an immediate implication: the hypothetical hadronic contribution must be approximately constant and universal in order to reproduce the observed NP‑like pattern.
The proposal that loops involving $D_{(s)}$ and $D^{(*)}$ resonances could account for the observed deviations, if not already included by the corresponding partonic computation, is therefore automatically disfavored. As shown in Refs.~\cite{Isidori:2024lng,Isidori:2025dkp}, monopole (dipole) photon couplings generate low- and high-$q^2$ contributions with opposite (the same) sign. Therefore, large effects at low-$q^2$, which require constructive monopole/dipole interference, correlate with much smaller effects at high-$q^2$, thus breaking the required universality. A constant and universal hadronic contribution remains a viable possibility.

At this point, a detailed examination of the theoretical  computation of the inclusive $B \to X_s\ell\ell$ decay in contrast with that of the  corresponding exclusive channels becomes essential. 
The crucial feature which we rely on is that, in the inclusive OPE, the effects of charm loops are under control and do not allow for the appearance of non-perturbative effects, like those described by the effective ${\cal C}_{9,\rm extra}$ coefficients we introduced above. Factorizable charm loops effects are {\it exactly} taken into account using $e^+e^- \to {\rm hadrons}$ data and the so-called Kr\"uger-Sehgal mechanism~\cite{Kruger:1996cv}. Nonfactorizable effects at low-$q^2$ are described, within SCET-I, in terms of $B$-meson subleading shape functions and are estimated conservatively at the 5\% level. The situation at high-$q^2$ is less clear cut. On one side, these nonfactorizable charm effects are given in terms of {\it local} $O(1/m_c^2)$ power corrections; on the other, the fact that the $q^2 > 15 \; {\rm GeV}^2$ region integrates over several broad charmonium resonances leaves open the possibility that some additional corrections might be present. Nonetheless, it is important to stress that, while these charm loop effects can, in principle, be modeled for inclusive decays, this is not the case for exclusive modes which rely on a large $q^2$ expansion.

In this work, motivated by the different sensitivities of inclusive and the corresponding exclusive decays to hadronic contributions, we introduce a new observable,  denoted by ${\cal F}$, that will, in the near future, allow one to discriminate between the two possibilities: genuine NP or a constant and universal hadronic contribution. Current data exhibit a certain preference for the NP interpretation, although the associated uncertainties are still substantial.
There is a large space for improvement especially on the inclusive mode and our projections show that in the high-$q^2$ region the ${\cal F}$ ratios can improve by more than 50\%. 

An important point that we have stressed is that all current PDG~\cite{ParticleDataGroup:2024cfk} and HFLAV~\cite{HeavyFlavorAveragingGroupHFLAV:2024ctg} averages for the various modes needed by LHCb and CMS to normalize their branching ratio measurements are currently calculated without taking properly into account the $\Upsilon \to (B^+B^-,B^0 \bar B^0)$ branching ratios. This effects have been discussed in the literature~\cite{Bernlochner:2023bad, Jung:2026ewj, HeavyFlavorAveragingGroupHFLAV:2024ctg} but they have not been implemented in the averages yet. Since the uncertainties on the various normalizing modes are already a large (and, in some cases, the largest) fraction  of the total uncertainties in LHCb and CMS branching ratios measurements, this issue has to be urgently resolved. In this work, we attempt to correct this problem using a two-pronged approach. First, using the fractions $f^{+-,00}$ determined in App.~\ref{ssec:BRcharmonium}, we recalculate the averages for all required charmonium branching ratios. Second, we remove the charmonium normalizations from the LHCb and CMS exclusive $b \to s \mu^+ \mu^-$ branching ratio measurements and reintroduce them using our updated averages.

Finally, at a more basic level, we compared the theoretical computation of the inclusive branching ratio with the one of the sum of exclusive branching ratios. By imposing the equivalence of the two computations, we found that a universal hadronic contribution is compatible 
with zero and remains rather small (in absolute value) compared to the preferred value provided by the global fits. Although current data still carry uncertainties that are too large to draw definite conclusions, the expected improvement in the precision of the inclusive mode could soon provide a more stringent constraint on this hypothetical universal hadronic contribution. We also showed that one can define a set of SM-normalized observables ($R^X$ with $X=\rm inc,\sum exc $) that enable a comparison of the ${\cal C}_9$ dependence of the inclusive and sum of exclusive modes with reduced sensitivity to input-parameter choices. This demonstrates, on the one hand, the excellent agreement between the ${\cal C}_9$ dependencies of the two computations and, on the other hand, again a preference for a NP interpretation over a universal hadronic contribution.

In summary we have shown that the imprint of a NP or a universal hadronic contribution in the observables once the inclusive mode is considered can be distinguished with a current preference for the NP solution.

\section*{Acknowledgements}

The authors are grateful to Kostas Petridis for helpful discussions.  J.M. gratefully acknowledges the financial support from ICREA under the ICREA Academia programme 2018 and to AGAUR under the Icrea Academia programme 2024 and from Departament de Recerca i Universitats de la Generalitat de Catalunya (SGR-00649). J.M. also received financial support from the Spanish Ministry of Science, Innovation and Universities (project PID2023-146142NB-I00) and from the IPPP Diva Award 2024. E.L. acknowledges support from CERN where part of this work was performed. The work of B.C. is supported by the La Caixa Junior Leader fellowship from the “la Caixa” Foundation (ID 100010434, fellowship code LCF/BQ/PI24/12040024).

\appendix

\section{Branching ratios of normalization modes}
\label{ssec:BRcharmonium}

In this appendix we calculate the experimental averages of the branching ratios for $B^{(*)+} \to K^{(*)+} J/\psi$, $B^{(*)0} \to K^{(*)0} J/\psi$ and $B^+ \to K^+ \psi(2S)$. Our starting point are the results collected in the PDG~\cite{ParticleDataGroup:2024cfk} or HFLAV~\cite{HeavyFlavorAveragingGroupHFLAV:2024ctg}.

A problem that affects almost all $B^{0,\pm}$ branching ratios measured at B-factories is the impact of $\Upsilon (4S)$ branching ratios to $B^+B^-$ ($f^{+-}$), $B^0\bar B^0$ ($f^{00}$) and modes without $b$-hadrons ($f_{\slash \!\!\!\! B}$), with $f^{+-} + f^{00} + f_{\slash \!\!\!\! B} = 1$. Currently almost all measured branching fractions which enter the PDG and HFLAV averages have been obtained with the $f^{+-} = f^{00} = 1/2$ assumption. 

This problem has been discussed recently in Refs.~\cite{Bernlochner:2023bad, Jung:2026ewj}. Here we follow the analysis presented in section 4.1 of the HFLAV review~\cite{HeavyFlavorAveragingGroupHFLAV:2024ctg} (which includes improvements suggested in Ref.~\cite{Bernlochner:2023bad}), and combine all existing information on $f^{+-}/f^{00}$, $f^{00}$ and $f_{\slash \!\!\!\! B}$ with a profile log-likelihood fit.

The only departure from the analysis presented in Ref.~\cite{HeavyFlavorAveragingGroupHFLAV:2024ctg} concerns the calculation of the $f^{+-}/f^{00}$ average which is then fed to the profile log-likelihood fit. First of all, we harmonize the values of the ratio of charged and neutral $B$ mesons used in the various measurements listed in Table 4 of Ref.~\cite{HeavyFlavorAveragingGroupHFLAV:2024ctg}. This has been done by dividing each measurement by the chosen ratio of lifetimes and multiplying the average by the latest PDG value $\tau(B^+)/\tau(B^0) = 1.075 \pm 0.004$~\cite{ParticleDataGroup:2024cfk}. A second improvement consists in the estimate of isospin breaking effects. The only $f^{+-}/f^{00}$ measurement which incorporates this estimate is the most recent Belle one~\cite{Belle:2022hka} where a 4.4\% isospin breaking uncertainty is added to the final result, following the analysis of isospin breaking corrections presented in Ref.~\cite{Fleischer:2001cw}. This uncertainty is 100\% correlated across all $f^{+-}/f^{00}$ measurements; therefore, we remove it from the Belle\cite{Belle:2022hka} result, average all measurements and reintroduce the uncertainty on the average. The result of this procedure is $(f^{+-}/f^{00})_{\rm average} = 1.063 \pm 0.050$. 

The $f^{+-}/f^{00}$ average has an uncertainty dominated by isospin breaking ($\pm 4.4\%$ from Ref.~\cite{Fleischer:2001cw}) and is considerably less precise than the result presented in Ref.~\cite{Jung:2026ewj} which reads $1.062 \pm 0.019$. We prefer to follow the procedure described above as it is a straightforward improvement on the HFLAV analysis and does not modify the estimate of the uncertainty associated with isospin breaking corrections presented in Refs.~\cite{Fleischer:2001cw, Belle:2022hka}.

\begin{figure}[t!]
    \centering
    \includegraphics[width=0.5\linewidth]{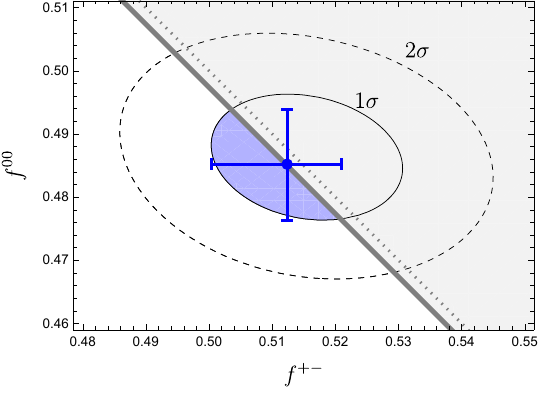} 
\caption{Result of the profile log-likelihood fit following the strategy presented in Ref.~\cite{HeavyFlavorAveragingGroupHFLAV:2024ctg}. The two contours correspond to $\chi^2_{\rm min} + 1$ and $\chi^2_{\rm min} + 4$. The gray region is excluded by a lower limit on $f_{\slash \!\!\!\! B}$ and causes the asymmetry in the fit result (in blue). }\label{fig:f+-f00}
\end{figure}

The profile log-likelihood fit is then identical to that described in Ref.~\cite{HeavyFlavorAveragingGroupHFLAV:2024ctg}. The two dimensional $\chi^2_{\rm min} + 1$ and $\chi^2_{\rm min} + 4$ contours that we obtain are presented in Fig.~\ref{fig:f+-f00}. After profiling over $f_{\slash \!\!\!\! B}$ we obtain the following symmetrized averages:
\begin{align}
    f^{+-} &= 0.511 \pm 0.010 \; , \label{eq:f+-} \\
    f^{00} &= 0.4851 \pm 0.0088 \; . \label{eq:f00}
\end{align}

\noindent For the $B\to K J/\psi$ modes, the PDG collects a series of measurements, all of which with the exception of the most recent Belle result assume $f^{+-} = f^{00} = 1/2$. We have verified that the PDG averages are simple weighted averages of these inputs. In order to properly include the effects of the measured $\Upsilon(4S)$ $b$-hadron fractions, we (1) remove them from the Belle results (using the $f^{+-}$ and $f^{00}$ adopted by Belle), (2) combine the latter with all other measurements, and (3) reintroduce the $b$-hadron fractions given in Eqs.~(\ref{eq:f+-}) and (\ref{eq:f00}) into the final averages. 

In the first step we remove the effect of $f^{+-(00)}\neq 1/2$ by multiplying by $2 f^{+-(00)}$ and reducing the uncertainty in quadrature\footnote{If the number of $B^+\to K^+J/\psi$ events is $N^+$ and the total number of $BB$ pairs is $N$ the inclusion of the ratio $f^{+-}$ causes the branching ratio to shift from ${\cal B}^{\rm 50:50}= N^+/(N/2)$ to ${\cal B}^{\rm correct} = N^+/(N f^{+-}) = {\cal B}^{\rm 50:50}/(2 f^{+-})$.}:
\begin{align}
    {\cal B} (B^+\to K^+ J/\psi)_{\rm Belle} & =
    \begin{cases}
        (10.32 \pm 0.25) \times 10^{-4} & \;\;[f^{+-} = 0.514 (6)] \\
        (10.61 \pm 0.23) \times 10^{-4} & \;\;[f^{+-} = 1/2] \\
    \end{cases},  \\
    {\cal B} (B^0\to K^0 J/\psi)_{\rm Belle} & =
    \begin{cases}
        (9.02 \pm 0.28) \times 10^{-4} & \;\;[f^{00} = 0.486 (6)] \\
        (8.77 \pm 0.25) \times 10^{-4} & \;\;[f^{00} = 1/2 ]\\
    \end{cases} .    
\end{align}
Then we average these branching ratios with all those included in the PDG average and reintroduce the fractions $f^{+-(00)}$ in the final average. The results are:\footnote{The PDG (HFLAV) averages in units of $10^{-4}$ for the $K^+$ and $K^0$ modes are $10.20 \pm 0.19$ ($10.18\pm 0.18$) and $8.91 \pm 0.21$ ($8.83 \pm 0.20$), respectively.}
\begin{align}
     {\cal B} (B^+\to K^+ J/\psi)_{\rm avg} & = (10.18 \pm 0.28) \times 10^{-4}, \label{eq:K+psi}\\
     {\cal B} (B^0\to K^0 J/\psi)_{\rm avg} & = (9.03 \pm 0.26) \times 10^{-4}. \label{eq:K0psi} 
\end{align}
The isospin averaged branching ratio we obtain from Eqs.~(\ref{eq:K+psi}) and (\ref{eq:K0psi}) is $9.57\; (19) \times 10^{-4}$.

For the $K^{*+(0)} J/\psi$ modes the situation is somewhat simpler because all measurements that enter the PDG average assume $f^{+-}=f^{00}=1/2$. Therefore we just need to rescale the PDG averages by $1/(2 f^{+-(00)})$:\footnote{The PDG (HFLAV) averages in units of $10^{-4}$ for the $K^{*+}$ and $K^{*0}$ modes are $14.30 \pm 0.81$ ($13.76\pm 0.82$) and $12.65 \pm 0.46$ ($12.78 \pm 0.44$), respectively.}
\begin{align}
       {\cal B} (B^+\to K^{*+} J/\psi)_{\rm avg} & = (13.97 \pm 0.85) \times 10^{-4}, \label{eq:K*+psi}\\
     {\cal B} (B^0\to K^{*0} J/\psi)_{\rm avg} & = (13.23 \pm 0.54) \times 10^{-4}. \label{eq:K*0psi} 
\end{align}
Finally we need the $B^+ \to K^+ \psi(2S)$ branching ratio. Here the situation is more complex because the PDG combination is not a simple average but is obtained via a more complex fit. Since, all measurements that enter the fit assume $f^{+-}=1/2$ we simply rescale their final result by $1/(2 f^{+-})$:
\begin{align}
    {\cal B} (B^+\to K^+ \psi(2S))_{\rm avg} & = (6.11 \pm 0.24) \times 10^{-4}. \label{eq:K+psi2S}
\end{align}

\section{Low-\texorpdfstring{$q^2$}{q2}}
In this appendix we calculate the averages for the normalized low-$q^2$ exclusive $b\to s\mu\mu$ branching ratios (Sec.~\ref{ssec:normalizedbsmmlow}) and then combine them with the normalizations calculated in App~\ref{ssec:BRcharmonium} (Sec.~\ref{ssec:bsmmlow}).

\subsection{\texorpdfstring{$B\to K^{(*)}\mu\mu$}{B -> K(*) mu mu} normalized rates}
\label{ssec:normalizedbsmmlow}

The four $B \to K^{(*)} \mu\mu$ decays have been measured for low and high-$q^2$ by LHCb~\cite{LHCb:2014cxe, LHCb:2016ykl}. The charged $B^+ \to K^+ \mu\mu$ mode has been measured also by CMS~\cite{CMS:2024syx}. Note that CMS presents results in narrow $1~{\rm GeV}^2$ bins which we have to combine paying attention not to dilute the uncertainty from the common $B^+ \to K^+ J/\psi$ normalization. We remove the normalizations adopted by LHCb and CMS and calculate the ratios 
\begin{align}
    {\cal R} (K^{+(0)} ) &\equiv {\cal B} (B^{+(0)} \to K^{+(0)} \mu\mu)/{\cal B} (B^{+(0)} \to K^{+(0)} J/\psi) ,\\
    {\cal R} (K^{*+(0)}) &\equiv {\cal B} (B^{+(0)} \to K^{*+(0)} \mu\mu)/{\cal B} (B^{+(0)} \to K^{*+(0)} J/\psi) .
\end{align}
For the $K^+$ case, we average LHCb and CMS. The results we obtain for the ratios ${\cal R} (K^{+(0)})$ are as follows. Note that, in order to remove the normalization, we need to use exactly what the experiments used.
\begin{itemize}
\item $B^+\to K^+ \mu\mu$ (LHCb~\cite{LHCb:2014cxe}, CMS~\cite{CMS:2024syx}):
\begin{align}
{\cal B} (B^+\to K^+ \mu\mu)_{\rm LHCb} &= (24.2 \pm 0.7 \pm 1.2 )\times (6-1.1)\times 10^{-9} = (1.186 \pm 0.068 )\times 10^{-7},  \\
{\cal B} (B^+\to K^+ J/\psi)_{\rm LHCb} &= (9.98 \pm 0.14 \pm 0.40) \times 10^{-4}=  (9.98 \pm 0.42 ) \times 10^{-4} , \\
{\cal R} (K^+)_{\rm LHCb} &= (1.188 \pm 0.046 ) \times 10^{-4}, \\\nonumber\\
{\cal B} (B^+\to K^+ \mu\mu)_{\rm CMS}  &= (1.253 \pm 0.057) \times 10^{-7}, \\
{\cal B} (B^+\to K^+ J\psi)_{\rm CMS}  &= (10.20 \pm 0.19) \times 10^{-4}, \\
{\cal R} (K^+)_{\rm CMS} &= (1.228 \pm 0.051 ) \times 10^{-4} , \\\nonumber\\
{\cal R} (K^+)_{\rm avg}^{\rm low} &= (1.206 \pm 0.034 ) \times 10^{-4} . \label{eq:RK+_low_avg}
\end{align}
\item $B^0\to K^0 \mu\mu$ (LHCb~\cite{LHCb:2014cxe}):
\begin{align}
{\cal B} (B^0\to K^0 \mu\mu)_{\rm LHCb} &= (18.85 \pm 3.35 \pm 0.9) \times (6-1.1) \times 10^{-9} =  (0.92 \pm 0.17 ) \times 10^{-7}, \\
{\cal B} (B^0\to K^0 J\psi)_{\rm LHCb} &= (9.28 \pm 0.13 \pm 0.37) \times 10^{-4} = (9.28 \pm 0.39)\times 10^{-4} , \\
{\cal R} (K^0)_{\rm avg}^{\rm low} &= (0.995 \pm 0.18 ) \times 10^{-4}.  \label{eq:RK0_low_avg}
\end{align}
\item $B^+\to K^{*+} \mu\mu$ (LHCb~\cite{LHCb:2014cxe}):
\begin{align}
{\cal B} (B^+\to K^{*+} \mu\mu) &= (36.95 \pm 7.95 \pm 2.6) \times (6-1.1) \times 10^{-9} =  (1.81 \pm 0.41 ) \times 10^{-7}, \\
{\cal B} (B^+\to K^{*+} J\psi) &= (1.431 \pm 0.027 \pm 0.090) \times 10^{-3} =  (1.431 \pm 0.094 ) \times 10^{-3}, \\
{\cal R} (K^{*+})_{\rm avg}^{\rm low} &= (1.265 \pm 0.27 ) \times 10^{-4}.  \label{eq:RK*+_low_avg}
\end{align}
\item $B^0\to K^{*0} \mu\mu$ (LHCb~\cite{LHCb:2016ykl}):
\begin{align}
{\cal B} (B^0\to K^{*0} \mu\mu) &= (0.342 \pm 0.017\pm 0.009 \pm 0.023) \times (6-1.1) \times 10^{-7} \nonumber\\
                                &=  (1.68 \pm 0.15 ) \times 10^{-7},\\
{\cal B} (B^0\to K^{*0} J\psi) &= (1.19 \pm 0.01 \pm 0.08) \times 10^{-3} =  (1.19 \pm 0.081 ) \times 10^{-3},\\
{\cal R} (K^{*0})_{\rm avg}^{\rm low} &= (1.408 \pm 0.078 ) \times 10^{-4}.  \label{eq:RK*0_low_avg}
\end{align}
\end{itemize}
Observe that the latest LHCb measurement of the $B^0 \to K^{*0} \mu\mu$ branching ratio~\cite{LHCb:2025mqb} adopts a normalization which is adjusted to the $m_{K^+\pi^-} \in [0.7459,1.0959]\; {\rm GeV}$ mass window. Since this quantity can only be reconstructed from the Belle~\cite{Belle:2014nuw} measurement, we refrain from calculating an updated result for the ratio ${\cal R}(K^{*0})$ and present directly the branching ratios given in Ref.~\cite{LHCb:2025mqb}.

\subsection{\texorpdfstring{$B\to K^{(*)}\mu\mu$}{B -> K(*) mu mu} branching ratios}
\label{ssec:bsmmlow}
Combining the results in App.~\ref{ssec:BRcharmonium} and Sec.~\ref{ssec:normalizedbsmmlow} we obtain the following branching ratios:
\begin{align}
{\cal B} (B^+ \to K^+ \mu\mu)_{\rm avg}^{\rm low} &= (1.228 \pm 0.048) \times 10^{-7} ,\\
{\cal B} (B^0 \to K^0 \mu\mu)_{\rm avg}^{\rm low} &= (0.90 \pm 0.16) \times 10^{-7} ,\\
{\cal B} (B^+ \to K^{*+} \mu\mu)_{\rm avg}^{\rm low} &= (1.77 \pm 0.40 )\times 10^{-7},\\
{\cal B} (B^0 \to K^{*0} \mu\mu)_{\rm avg}^{\rm low} &= (1.578 \pm 0.081 )\times 10^{-7},
\end{align}
where the last branching ratio is taken directly from Ref.~\cite{LHCb:2025mqb}. 

The isospin average rates, which are the quantities that appear in the ratios ${\cal F}_{K^{(*)}}^{[1.1,6]}$ are:
\begin{align}
{\cal B} (B \to K \mu\mu)_{\rm avg}^{\rm low} &=(1.202 \pm 0.046) \times 10^{-7} , \label{eq:BRK_low}\\
{\cal B} (B \to K^{*} \mu\mu)_{\rm avg}^{\rm low} &= (1.585 \pm 0.079 )\times 10^{-7}. \label{eq:BRK*_low}
\end{align}
Note that the isospin averaged branching ratios do not assume the isospin equality of charged and neutral $B\to K^{(*)} \mu\mu$ partial widths. Moreover, we have explicitly included an estimate of isospin breaking effects in the $f^{+-}$ and $f^{00}$ determinations. 

\section{High-\texorpdfstring{$q^2$}{q2}}
In this appendix we calculate the averages for the normalized high-$q^2$ exclusive $b\to s\mu\mu$ branching ratios (Sec.~\ref{ssec:normalizedbsmmhigh}) and then combine them with the normalizations calculated in App.~\ref{ssec:BRcharmonium} (Sec.~\ref{ssec:bsmmhigh}).

\subsection{\texorpdfstring{$B\to K^{(*)}\mu\mu$}{B -> K(*) mu mu} normalized rates}
\label{ssec:normalizedbsmmhigh}

We follow the exact same procedure as in Sec.~\ref{ssec:normalizedbsmmlow} to calculate the corresponding ${\cal R}(K^{*})$ for $q^2 > 15\; {\rm GeV}^2$.  
\begin{itemize}
\item $B^+\to K^+ \mu\mu$ (LHCb~\cite{LHCb:2014cxe}, CMS~\cite{CMS:2024syx}):
\begin{align}
{\cal B} (B^+\to K^+ \mu\mu) &= (12.1 \pm 0.4 \pm 0.6 ) \times (22-15) \times 10^{-9} =  (0.847 \pm 0.050 ) \times 10^{-7}, \\
{\cal B} (B^+\to K^+ J\psi) &= (9.98 \pm 0.14 \pm 0.40) \times 10^{-4} , \\
{\cal R} (K^+)_{\rm LHCb} &= (0.849 \pm 0.035 ) \times 10^{-4},   \\\nonumber\\
{\cal B} (B^+\to K^+ \mu\mu) &= (0.927 \pm 0.046) \times 10^{-7}, \\
{\cal B} (B^+\to K^+ J\psi) &= (10.20 \pm 0.19) \times 10^{-4}, \\
{\cal R} (K^+)_{\rm CMS} &= (0.909 \pm 0.042 ) \times 10^{-4},\\\nonumber\\
{\cal R} (K^+)_{\rm avg}^{\rm high} &= (0.874 \pm 0.027 ) \times 10^{-4} . \label{eq:RK+_high_avg}
\end{align}
\item $B^0\to K^0 \mu\mu$ (LHCb~\cite{LHCb:2014cxe}):
\begin{align}
{\cal B} (B^0\to K^0 \mu\mu) &= (9.55 \pm 1.55 \pm 0.5) \times (22-15) \times 10^{-9} =  (0.67 \pm 0.11 ) \times 10^{-7}, \\
{\cal B} (B^0\to K^0 J\psi) &= (9.28 \pm 0.13 \pm 0.37) \times 10^{-4} =  (9.28 \pm 0.39 ) \times 10^{-4}, \\
{\cal R} (K^0)_{\rm avg}^{\rm high} &= (0.72 \pm 0.12 ) \times 10^{-4}.  \label{eq:RK0_high_avg}
\end{align}
\item $B^+\to K^{*+} \mu\mu$ (LHCb~\cite{LHCb:2014cxe}):
\begin{align}
{\cal B} (B^+\to K^{*+} \mu\mu) &= (39.85 \pm 7.65 \pm 2.8) \times (19-15) \times 10^{-9} \nonumber\\
                                &=  (1.59 \pm 0.33 ) \times 10^{-7}, \\
{\cal B} (B^+\to K^{*+} J\psi) &= (14.31 \pm 0.27 \pm 0.90) \times 10^{-4} =  (14.31 \pm 0.94 ) \times 10^{-4}, \\
{\cal R} (K^{*+})_{\rm avg}^{\rm high} &= (1.11 \pm 0.22 ) \times 10^{-4}.  \label{eq:RK*+_high_avg}
\end{align}
\item $B^0\to K^{*0} \mu\mu$ (LHCb~\cite{LHCb:2016ykl}):
\begin{align}
{\cal B} (B^0\to K^{*0} \mu\mu) &= (0.4355 \pm 0.0185\pm 0.007 \pm 0.030) \times (19-15) \times 10^{-7}\nonumber\\
                                &=  (1.74 \pm 0.14 ) \times 10^{-7},\\
{\cal B} (B^0\to K^{*0} J\psi) &= (1.19 \pm 0.01 \pm 0.08) \times 10^{-3} =  (1.19 \pm 0.081 ) \times 10^{-3}, \label{eq:BRK*0psiOriginal}\\
{\cal R} (K^{*0})_{\rm avg}^{\rm high} &= (1.464 \pm 0.069 ) \times 10^{-4}.  \label{eq:RK*0_high_avg}
\end{align}
Similarly to the low-$q^2$ case, the latest LHCb measurement of the $B^0 \to K^{*0} \mu\mu$ branching ratio~\cite{LHCb:2025mqb} adopts a normalization which is adjusted to the $m_{K^+\pi^-} \in [0.7459,1.0959]\; {\rm GeV}$ mass window. We refrain from calculating an updated result for the ratio ${\cal R}(K^{*0})$ and present directly the branching ratios given in Ref.~\cite{LHCb:2025mqb}. 
\item $B^+\to K^+\pi^+\pi^- \mu\mu$ (LHCb~\cite{LHCb:2014osj}):
\begin{align}
{\cal B} (B^+\to K^+\pi^+\pi^- \mu\mu) &= (19-15) ( 0.11 \pm 0.07 \pm 0.01 ) \times 10^{-8} = ( 0.44 \pm 0.28) \times 10^{-8} , \\
{\cal B} (B^+\to K^{+} \psi(2S)) &= (6.27 \pm 0.24 ) \times 10^{-4} ,\\
{\cal R} (K^+\pi^+\pi^-) &= ( 0.70 \pm 0.45 ) \times 10^{-5} . \label{eq:RKpipi_high_avg}
\end{align}
\end{itemize}

\subsection{\texorpdfstring{$B\to K^{(*)}\mu\mu$}{B -> K(*) mu mu} branching ratios}
\label{ssec:bsmmhigh}
Combining the results in App.~\ref{ssec:BRcharmonium} and Sec.~\ref{ssec:normalizedbsmmhigh} we obtain the following branching ratios:
\begin{align}
{\cal B} (B^+ \to K^+ \mu\mu)_{\rm avg}^{\rm high} &= (0.890 \pm 0.036) \times 10^{-7} ,\\
{\cal B} (B^0 \to K^0 \mu\mu)_{\rm avg}^{\rm high} &= (0.65 \pm 0.11) \times 10^{-7} ,\\
{\cal B} (B \to K \mu\mu)_{\rm avg}^{\rm high} &= (0.866 \pm 0.035) \times 10^{-7} , \label{eq:BRK_high}\\
{\cal B} (B^+ \to K^{*+} \mu\mu)_{\rm avg}^{\rm high} &= (1.56 \pm 0.32 )\times 10^{-7},\\
{\cal B} (B^0 \to K^{*0} \mu\mu)_{\rm avg}^{\rm high} &= (1.780 \pm 0.091 )\times 10^{-7},\\
{\cal B} (B \to K^{*} \mu\mu)_{\rm avg}^{\rm high} &= (1.763 \pm 0.088 )\times 10^{-7}, \label{eq:BRK*_high}\\
{\cal B} (B^- \to (K\pi)_S \mu\mu)_{\rm avg}^{\rm high} &= (0.11 \pm 0.04)\times 10^{-7}, \label{eq:BRKpiS}\\
{\cal B} (B^- \to K^+\pi^+\pi^- \mu\mu)_{\rm avg}^{\rm high} &= (0.43 \pm 0.28)\times 10^{-8},
\end{align}
where the $K^*$ and $(K^+\pi^-)_S$ branching ratios are taken directly from Ref.~\cite{LHCb:2025mqb}. Notice that the $S$-wave branching ratio in Eq.~(\ref{eq:BRKpiS}) contains a $3/2$ isospin factor required to take into account the $K_S\pi^0$ mode. 

Moreover, assuming isospin invariance of the $B^{+,0}\to (K\pi)_S \mu\mu$ partial widths,  we can write the following isospin average:
\begin{align}
    {\cal B}(B\to (K\pi)_S \mu\mu) &=  {\cal B}(B^0\to (K\pi)_S \mu\mu) \frac{\tau^0 + \tau^+}{2 \tau_0} \\
    &= ( 0.114 \pm 0.042 ) \times 10^{-7} \label{eq:BRKpiS_high}
\end{align}
Finally, using isospin once more we can obtain an expression for the isospin averaged $B\to K\pi\pi \mu\mu$ branching ratio:
\begin{align}
{\cal B} (B \to K\pi\pi \mu\mu) & = \frac{{\cal B} (B^+\to K\pi\pi \mu\mu)}{\tau^+} \frac{\tau^0 + \tau^+}{2} \\
&=\frac{3}{2} \frac{\tau^0 + \tau^+}{2\tau^+}{\cal B} (B^+\to K^+\pi^+\pi^- \mu\mu) \label{eq:Kpipimid}\\
&= ( 0.062 \pm 0.040 ) \times 10^{-7}, \label{eq:BRKpipi_high}
\end{align}
where we used $\tau^0/\tau^+ = 1.076 \pm 0.004$~\cite{ParticleDataGroup:2024cfk}. The explicit factor of $3/2$ in Eq.~(\ref{eq:Kpipimid}) takes into account modes involving neutral $K$ and $\pi$ (see the discussion in section 3.2 of Ref.~\cite{Huber:2024rbw}).

\bibliographystyle{JHEP}
\bibliography{references}

\end{document}